# How the "Auction Cube" Supports the Selection of Auction Designs in Industrial Procurement

# Wahl des Auktionsverfahrens im industriellen Einkauf mit Hilfe des „Auktionswürfels"

- Gregor Berz, Institut für Angewandtes Mechanism Design, gregor.berz@ifamd.de
- Florian Rupp, Institut für Angewandtes Mechanism Design, Kutaisi International University, florian.rupp@ifamd.de
- Brian Sieben, Hilti Aktiengesellschaft, brian.sieben@hilti.com

**Abstract:**

Purpose:

It is well known that rightly applied reverse auctions offer big commercial potential to procurement departments. However, the sheer number of auction types often overwhelms users in practice. And since the implications of a wrongly chosen auction type are equally well known, the overall usage of reverse auctions lacks its potential significantly. In this paper, a novel method is being proposed that guides the user in selecting the right combination of basic auction forms for single lot events, considering both market- as well as supplier-related, bijective criteria.

Approach:

Assuming a Gaussian distribution for bidders, their strategic margin is benchmarked with a threshold derived from the significance Sigma of the bidder's identic distribution. This delivers a basic criterion for the trade off between first price and second price auctions. Following Milgrom`s seminal approach, the revenue equivalence theorem is interpreted reversely in order to derive more differentiated recommendations.

Findings:

In this paper we use that the mean of order statistics of any sample size n of the Gaussian distribution with significance Sigma is proportional in Sigma. In the appendix of this paper the constant values $g(n)$ and $h(n)$ are listed. These values $g(n)$ directly support the decision whether a first price auction is expected to deliver better results than a second price auction. These values $h(n)$ reveal whether a Vickrey Auction is most likely leading to better results than a Hongkong Auction.

Research limitations/implications:

The thresholds presented in this paper for the decision between certain auction types can be understood as exact values only as precise as $g(n)$ and $h(n)$ respectively the underlying expectation values of order statistics of normal distributions with Sigma = 1 are calculated. We calculate here with 10 significant digits.

Practical implications:

The proposed method allows for a quick and robust auction design, leveraging the tremendous commercial potential of reverse auctions.

Originality/value:

To the best of the authors` knowledge, this is the first paper that proposes a robust method to guide practitioners in selecting the right combination of basic auction types considering market and commercial criteria. Especially noteworthy is that the paper underpins the method by means of order statistics.

**Einleitung**

Seit Anfang der 2000er Jahre sind Reverse Auktionen im industriellen Einkauf etabliert. Bei Reversen Auktionen bieten Lieferanten auf Aufträge des Kunden, die von dessen Einkauf vergeben werden. Die Wahl des angemessenen Auktionsverfahrens je nach vorhandener Markt- und Wettbewerbssituation stellt allerdings bis heute eine Herausforderung für den Einkäufer dar. Die Theorie bietet Basisauktionsformen an, die in Ausschlussrunden kombiniert werden können. Die vorliegende Veröffentlichung leitet anhand der drei Kriterien ‚Anzahl der Bieter', ‚Streuung der Angebote' und ‚Risikoaversion der Bieter' für Entscheidungssituationen mit einem Verhandlungsobjekt jeweils ein eindeutiges, aus den Basisauktionsformen kombiniertes Auktionsverfahren her. Wir folgen dabei der Anleitung von Paul Milgrom, der sagte: „One important use of the revenue equivalence theorem is as a benchmark for analysing cases when the assumptions of the theorem do not hold" ([Milgrom 2004], Seite 77) und interpretieren das Erlösäquivalenztheorem der Auktionstheorie rückwärts.

Als Basisauktionsformen betrachten wir die wesentlichen, in industriellen Industriegütermärkten relevanten Formen ([Berz 2014] Seiten 44 und 59): verdeckte Erstpreisauktion, verdeckte Zweitpreisauktion, Englische Auktion (als Ticker oder als Dynamische Englische Auktion), Hongkong Auktion und Holländische Auktion.

Im ersten Teil dieser Veröffentlichung vergleichen wir Erst- und Zweitpreisauktionen. Für die Indifferenzpreise der Bieter wird die Normalverteilung NV(0, Sigma) mit Erwartungswert Null und Standardabweichung Sigma angenommen. Für die, aufgrund der Risikoaversion zu erwartende, Strategische Marge M der Bieter wird eine Grenze in Abhängigkeit von Sigma und der Anzahl der Bieter n angegeben, bis zu der eine Erstpreisauktion den niedrigeren Erwartungswert liefert. Liegt die Strategische Marge über dieser Grenze, ist der Erwartungswert einer Zweitpreisauktion besser. Die Grenze wird im ersten Schritt mit 2 * sqrt(3) * Sigma /(n+1) angenähert. Im zweiten Schritt betrachten wir den relativen Fehler dieser Näherung. Bei den exakten Berechnungen für 28 Ausprägungen von Sigma beobachten wir, jeweils bei fester Anzahl der Bieter n, einen identischen relativen Fehler. Der relative Fehler dieser Näherung ist tatsächlich unabhängig von Sigma, weil schon der Erwartungswert $E(X,i)$ der i-ten

Ordnungsstatistik der Normalverteilung NV(0, Sigma) mit fester Stichprobenlänge n immer in Sigma proportional ist.. D.h., für jedes n gibt es einen Faktor g(n), so dass Sigma * g(n) die exakte Grenze darstellt. Wir geben im Anhang die Faktoren g(n) für alle ganzzahligen n zwischen 2 und 100 auf 10 Stellen genau an.

Im zweiten Teil der Veröffentlichung diskutieren wir Kriterien für die Frage, ob in einer vorliegenden Entscheidungssituation die Veranstaltung einer Auktion generell als empfehlenswert einzustufen ist oder nicht. Dabei gehen wir auch auf die Gefahren kollusiven Bieterverhaltens und auf asymmetrische Präferenzen des Auktionators in Bezug auf die Bieter ein. Schließlich formulieren wir drei Voraussetzungen, die erfüllt sein müssen, damit der in Teil 4 der Veröffentlichung vorgestellte „Auktionswürfel" anwendbar ist.

Im dritten Teil der Veröffentlichung diskutieren wir die erwarteten Ergebnisse der Basisauktionsformen in Abhängigkeit von Sigma, n und M. Ferner analysieren wir Eigenschaften dieser Auktionsformen, die sich als Erfahrungswerte aus der industriellen Praxis ergeben. Dabei gehen wir auch auf Maßnahmen gegen kollusives Bieterverhalten und auf den „Fluch des Gewinners" ein.

Im vierten Teil leiten wir unter der Annahme, dass die Kriterien aus Teil 2 für die Veranstaltung einer Auktion sprechen, für jede der acht Konstellationen (n klein, n groß) x (M klein, M groß) x (Sigma klein, Sigma groß) - das sind die „Ecken des Auktionswürfels" – ein kombiniertes Auktionsverfahren her, das in der Praxis als empfehlenswert einzustufen ist.

Weitere Stichwörter (die in der Einleitung nicht vorkommen, da Synonyme verwendet werden): Auktionsdesign, Auktion mit einem Los, Japanische Auktion, Einkaufsauktion

**Teil 1: Abwägung zwischen Erst- und Zweitpreisauktion.**

Wir betrachten eine Entscheidungssituation mit n Bietern, n > 1, deren Indifferenzpreise für das Verhandlungsobjekt identisch normalverteilt seien mit Standardabweichung Sigma. Der Erwartungswert kann ohne Beschränkung der Allgemeingültigkeit als 0 angenommen werden. Als Indifferenzpreis bezeichnen wir diejenige Bewertung des Verhandlungsobjekts, die einen Bieter indifferent sein lässt darin, ob er das Objekt verkauft oder nicht. Der Truth-revealing-Effekt einer Zweitpreisauktion - dass nämlich ein Bieter mit seinem Gebot nur beeinflussen kann ob er gewinnt, nicht aber den Preis beeinflussen kann, den er erhält, wenn er gewinnt – sorgt bekanntlich (wie wir heute wissen, wusste schon Goethe davon, siehe [Holler 2020] Seite 239) dafür, dass der Indifferenzpreises als Angebot die streng dominante Angebotsstrategie für alle Bieter ist. Damit ist das erwartete Ergebnis einer Zweitpreisauktion E(SPSB) (SPSB für „Second Price Sealed Bid") der Erwartungswert der zweiten Ordnungsstatistik X(2,n) mit Stichprobenlänge n der Normalverteilung N(0, Sigma). E(SPSB) = E(X(2,n)).

Für die Angebotsstrategie in einer Erstpreisauktion muss für dieselben Bieter angenommen werden, dass keiner seinen Indifferenzpreis bietet. Denn würde ein Bieter seinen Indifferenzpreis bieten und damit den Zuschlag der Auktion erhalten, hätte er - per Definition des Indifferenzpreises – keinen Gewinn. Also wird jeder Bieter in einer Erstpreisauktion eine Strategische Marge ([Berz 2014] Seite 80f) auf seinen

Indifferenzpreis addieren um sein Angebot zu berechnen. Die Höhe dieser Strategischen Marge M ist reziprok zu seiner Aversion gegen das Risiko, den Zuschlag nicht zu bekommen. Damit ist das erwartete Ergebnis einer Erstpreisauktion E(FPSB) (FPSB für „First Price Sealed Bid") die Summe von M und dem Erwartungswert der ersten Ordnungsstatistik X(1,n) mit Stichprobenlänge n der Normalverteilung NV(0,Sigma). E(FPSB) = E(X(2,n)) + M.

Eine Erstpreisauktion liefert also genau dann im Erwartungswert ein besseres Ergebnis für den Auktionator als eine Zweitpreisauktion, wenn M < E(X(2,n)) – E(X(1,n)) =: G gilt.

Leider lassen sich die Erwartungswerte der Ordnungsstatistiken der Normalverteilung nicht exakt bestimmen. Als Integrale von Exponentialfunktionen lassen sich wohl konkrete Werte beliebig genau berechnen, es gibt aber keine geschlossene Formel in Sigma und n für diese Werte. Vermutlich ist das ein Grund, warum in der Spieltheorie-Literatur oft Gleichverteilungen für die Bewertungen von Auktionsobjekten seitens der Bieter angenommen werden. Die i-ten Ordnungsstatistiken Y(i,n) der Gleichverteilung auf [0,1] sind die Beta-Verteilungen Beta(i, n-i+1) mit den Erwartungswerten E(Y(i,n)) = i/(n+1) ([Kabluchko 2017] Beispiel 1.6.4 und Aufgabe 1.6.6). Mit diesen Formeln lässt sich sehr einfach rechnen und argumentieren.

Die Gleichverteilungsannahme hat allerdings einen großen Kritikpunkt: Auch die „stetige" Gleichverteilung auf (0,1) ist spätestens an den Rändern des zugrundeliegenden Intervalls unstetig und wird damit zu einer in der Praxis nicht beobachteten Wahrscheinlichkeitsverteilung. Die Frage, welche Werte als obere und untere Intervallgrenzen der Gleichverteilung angenommen werden sollen, kann in der Praxis regelmäßig nicht beantwortet werden. Aus dem abgeschlossenen Intervall [0,1] das offene Intervall (0,1) zu machen um die „Stetigkeit" zu retten ist ein mathematischer Kunstgriff und letztlich nur eine kosmetische Reparatur. Es können keineswegs „ohne Beschränkung der Allgemeingültigkeit" die Intervallgrenzen 0 und 1 angenommen werden, denn die Gleichverteilung auf (0,1) hat immer die gleiche Standardabweichung 1/(2*sqrt(3)). In der Praxis sind aber Bieter mit unterschiedlichen Streuungen ihrer Indifferenzpreise zu beobachten. Dies ist mit der in der Auktionstheorie häufigen Normierungsannahme auf das kompakte Intervall (0,1) nicht darstellbar.

Um nun die Erwartungswerte der 1. und der 2. Ordnungsstatistik der Normalverteilung anzunähern, bedienen wir uns der Annahme, dass sich die Ordnungsstatistiken Y' einer Gleichverteilung mit Erwartungswert 0 und Standardabweichung Sigma ähnlich verhalten wie diejenigen der Normalverteilung mit gleichem Erwartungswert und Sigma. Für die Ordnungsstatistiken Y'' der Gleichverteilung auf (-1/2, 1/2) mit Standardabweichung Sigma''=1/(2*sqrt(3)) erhalten wir E(Y''(1,n))= 1/(n+1) - 1/2 = (1-n)/(2n+2) und E(Y''(2,n))= 2/(n+1) - 1/2 = (3-n)/(2n+2). Nach Skalierung mit 2*sqrt(3)*Sigma zur Gleichverteilung auf (-sqrt(3) *Sigma, sqrt(3) *Sigma) mit Standardabweichung Sigma folgt E(Y'(1,n))= 2*sqrt(3)*Sigma*(1-n)/(2n+2) und E(Y'(2,n))= 2*sqrt(3)*Sigma*(3-n)/(2n+2).

Damit können wir in erster Näherung eine Grenze G' für M angeben, unterhalb der eine Erstpreisauktion ein für den Auktionator besseres Ergebnis im Erwartungswert liefert als eine Zweitpreisauktion: M < E(Y'(2,n)) – E(Y'(1,n)) = 2*sqrt(3)*Sigma/(n+1) =: G'.

Damit kann das Erlösäquivalenztheorem ([Milgrom 2004] Abschnitt 3.3.4 oder auch [Klemperer 2004] Abschnitt 1.4) – ein Kernstück der Auktionstheorie – rückwärts interpretiert werden. Das Erlösäquivalenztheorem geht neben anderen Voraussetzungen von risikoneutralen Bietern aus mit Indifferenzpreisen, die einer identischen stetigen Wahrscheinlichkeitsverteilung entsprechen. Das Erlösäquivalenztheorem besagt also - rückwärts interpretiert - , dass unter den unterstellten Voraussetzungen von einem risikoneutralen Bieter exakt die Strategische Marge M = G' = 2*sqrt(3)*Sigma/(n+1) zu erwarten ist. Oder, wenn die Gleichverteilung auf dem Intervall (0,1) angenommen wird, wie es in der Auktionstheorieliteratur üblicherweise normiert wird, dann ist Sigma = 1/(2*sqrt(3)) und damit M = 1/(n+1) die zu erwartende Strategische Marge eines risikoneutralen Bieters. Dies ist auch genau die Differenz der Erwartungswerte der ersten und der zweiten Ordnungsstatistik auf der Gleichverteilung auf (0,1). Um dieses Ergebnis aber in der Praxis auf Bieter mit beobachtetem beliebigem Sigma anzuwenden, ist das Intervall (a,b), auf dem die Indifferenzpreise als gleichverteilt anzunehmen sind, geeignet zu wählen so dass die Standardabweichung (b-a)/(2*sqrt(3)) = Sigma ist - das entspricht genau dem oben beschriebenen Ansatz. Die Ausprägung von Sigma ist in der gewählten Intervallbreite b-a der zu unterstellenden Gleichverteilung inhärent.

Betrachtet man – um eine randlos stetige Wahrscheinlichkeitsverteilung zu unterstellen - die Indifferenzpreise der Bieter als normalverteilt mit Standardabweichung Sigma, dann interessiert natürlich noch der Fehler, der in der Abschätzung durch G' = 2*sqrt(3)*Sigma/(n+1) als Grenze für M liegt. Um diesen Fehler einzuschätzen haben wir die exakten Werte für G bzw. für E(X(i,n)) für i = 1, 2 jeweils für alle n = 2, 3, 4, 5 und 6 sowie für alle Sigma = 0.1, 0.2, ..., 1, 2, ..., 10, 20, ..., 100 mit dem Computeralgebrasystem Maple auf zehn signifikante Stellen genau berechnet. Das Ergebnis findet sich im Anhang. Die Beobachtung ist, dass jeweils für alle n der relative Fehler (G'-G)/G für alle Sigma den exakt selben Wert hat. Tatsächlich ist schon E(X(i,n)) proportional in Sigma, wie sich leicht aus der Skalierungseigenschaft der Normalverteilung N(0,Sigma) = N(0,1) * Sigma schließen lässt[1]. Damit existiert für jedes n ein konstanter Wert g(n) mit dem unabhängig von der Standardabweichung Sigma für die Grenze G der Strategischen Marge M gilt: G = E(X(2,n))- E(X(1,n)) = Sigma * g(n). Im Anhang finden sich alle diese Faktoren g(n) = G / Sigma für 1 < n < 101, jeweils berechnet mit Sigma = 1.

Beispiele:

Nehmen wir an, vier Bieter haben in einer ersten Angebotsrunde die Preise P1 = 112,18 EUR, P2 = 95,75 EUR, P3 = 109,27 EUR und P4 = 94,15 EUR abgegeben. Dann ist die beste Schätzung für Sigma, die wir zu diesem Zeitpunkt haben, die Stichprobenvarianz Quadratwurzel(Summe((Pi - Mittelwert(Pi, i=1, .., 4))^2, i=1, ..., 4)/n) = 7,97440711.

---

[1] Betrachtet man die Dichtefunktion der Normalverteilung und deren Integrale, dann mag der Zusammenhang überraschend wirken. Man kann aber die Realisierung der Normalverteilung N(0,Sigma) auch als Multiplikation der Zufallswerte einer Realisierung der Normalverteilung N(0,1) mit Sigma verstehen. Damit ist unmittelbar klar, dass sich die Skalierung mit Sigma auch durch die Realisierungen der Ordnungsstatistiken und deren Erwartungswerte drückt. Wir danken Stefan Napel für diesen Hinweis.

Für G' ergibt sich G' = 2*sqrt(3)*Sigma/(n+1) = 5,52483131. Unterstellt man den Bietern also Strategische Margen die kleiner sind als G'=5,52 EUR, wird eine Erstpreisauktion im Erwartungswert dem Auktionator bessere Ergebnisse liefern. Andernfalls ist eine Zweitpreisauktion vorzuziehen.

Unter der wohl praxisnäheren Annahme, dass die Bewertungen der Bieter normalverteilt sind, nehmen wir den Wert g(4) = 0,732363991 aus dem Anhang um G zu berechnen: G = g(n)*Sigma = 5,840168614. Bei dieser Betrachtung können den Bietern Strategische Margen bis G=5,84 EUR unterstellt werden, um ein für den Auktionator besseres Ergebnis von einer Erstpreisauktion zu erwarten als von einer Zweitpreisauktion.

Nehmen wir an, es hat noch ein fünfter Bieter einen Preis P5 = 97,33 EUR abgegeben. Dann ergibt sich eine neue Schätzung für Sigma = 7,200248385 und ein neues G' = 4,157065343 EUR. Für G verwenden wir g(5) = 0,667945504 und berechnen: G = 4,809373533 EUR.

Nehmen wir in einer anderen Situation an, vier Bieter haben in einer ersten Angebotsrunde die Preise P1 = 100,73 EUR, P2 = 100,86 EUR, P3 = 99,05 EUR und P4 = 100,12 EUR abgegeben. Dann erhalten wir als Schätzung für Sigma die Stichprobenvarianz 0,715017482. In diesem Beispiel ergibt sich G' = 0,495378643 EUR und G = 0,523653057 EUR.

Die unterstellte Strategische Marge der Bieter, die noch zugelassen ist, um eine Erstpreisauktion gegenüber einer Zweitpreisauktion vorzuziehen, hängt also signifikant von Sigma ab. Für Sigma liefert die Stichprobenvarianz vor einer Auktion bereits verfügbarer Preise derselben Bieter eine sehr grobe Näherung. Für die zu erwartenden Strategischen Margen sind diese verfügbaren Preise mit einer Strukturkostenanalyse zu vergleichen, die einen eigenen Einblick in das Niveau der Indifferenzpreise geben kann.

**Teil 2: Voraussetzungen für Auktionen in der Praxis**

Als wichtigster Störfaktor gegen Auktionen wird in der Praxis häufig eine Koordination unter den Bietern beobachtet, die ohne explizite Absprachen stattfindet und deshalb nicht wettbewerbsrechtlich relevant ist. Bieter, die lieber eine etablierte Marge im Markt bewahren wollen als sich gegenseitig den Preis zu reduzieren bieten weniger aggressiv und überlassen im Zweifel einzelnes Geschäft einem Wettbewerber. In der Spieltheorieliteratur wird dieses strategische Verhalten auch „Strategische Angebotsreduzierung" genannt, da es meist mit reduzierten, eigenen angebotenen Mengen einhergeht (z.B. [Klemperer 2004] Seite 64). Wir nennen solche Märkte kollusive Märkte. Märkte sind tendenziell dann kollusiv, wenn zwar in Summe aller Mengen ein Angebotsüberhang besteht, dabei aber wenige Anbieter vielen Abnehmern gegenüberstehen. Dann schaffen die wenigen Anbieter durch ihre Koordination – oft eben auch ohne explizite Absprache; es genügt die Transparenz der Preise z.B. mittels eines öffentlichen Index zur Koordination – die Verhandlungsmacht im Markt zu drehen und den Markt, der aufgrund des rein mengenmäßig gegebenen Angebotsüberhangs ein Einkäufermarkt sein müsste, auszuhebeln und einen Verkäufermarkt zu „simulieren". In einer solchen Marktkonstellation liefert die durch eine Auktion erhöhte Preistransparenz unter den Bietern eher noch eine Verstärkung der Kollusion, weshalb hier eine Auktion ggf. zu einem schlechteren Ergebnis führt als andere, auf die Kollusion zugeschnittene Verhandlungs- und Entscheidungsmechanismen. Als Voraussetzung für die Empfehlung, eine Auktion durchzuführen, formulieren wir hier ganz lapidar „Kein kollusiver Markt", wobei eine genauere Untersuchung der Einflussfaktoren

- Anzahl der Anbieter im Markt
- Anzahl der Nachfrager im Markt
- Häufigkeit von Lieferantenentscheidungen im Markt
- allgemeine Preistransparenz im Markt
- Ausprägung von Markteintrittsbarrieren

und ggf. noch weitere, z.B. auf die historischen Beziehungen zwischen den Marktbegleitern abzielende Kriterien eine eigene Untersuchung wert sind.

Die zweite Voraussetzung dafür, dass eine Auktion das beste zu erwartende Ergebnis in einer Lieferanten-Entscheidungssituationen für den Einkauf des Kunden liefert ist die absolute Preissensitivität des Kunden. Nur wenn jeder minimale Preisunterschied zwischen zwei Alterativen bereits dazu führt, dass der Kunde besser gestellt ist wenn er die günstigere Alternative wählt, wird eine Auktion genau diese Präferenz optimieren. Um dies zu erreichen hat sich in der industriellen Praxis seit vielen Jahren die Methode der ‚Bonussysteme' etabliert ([Berz 2014] Seite 135ff), die als monetäre Bewertung aller entscheidungsrelevanter Unterschiede zwischen den Bietern und deren Angeboten – also sowohl alle Vertragseigenschaften das konkrete Verhandlungsobjekt betreffend als auch Unternehmenseigenschaften des Bieters als potenziellen Vertragspartner - eine Preisdiskriminierung darstellt, mit der die Preissensitivität des Kunden maximiert ist.

Wenn wir als zweite Voraussetzung für die Empfehlung, eine Auktion durchzuführen, ganz lapidar „vergleichbare Bieter" formulieren, dann meinen wir damit zum einen, dass alle Bewertungsunterscheide seitens des Auktionators für die Bieter durch diese Preisdiskriminierung abgedeckt werden und der Kunde danach tatsächlich absolut preissensitiv ist. Gleichzeitig wird mit „vergleichbare Bieter" auch gefordert, dass allen Bietern die identische Risikoaversion dagegen, das Geschäft nicht zu gewinnen, unterstellt werden kann. Dies ist notwendig, um in der Abwägung zwischen Erst- und Zweitpreisauktion eine für alle Bieter einheitliche Strategische Marge M unterstellen zu können.

Schließlich sind noch Entscheidungssituationen zu erwähnen, in welchen interne strategische Gründe dagegen sprechen, einen Lieferanten der als Vertrauenspartner am meisten Mehrwert generiert, überhaupt Wettbewerb auszusetzen. In diesem Fall ist die kooperative Spieltheorie aufzurufen, um die Mitte des mit dem Vertrauenspartner gemeinsam generierten Kuchens zu finden. Für die hier betrachtete Fragestellung, ob die Veranstaltung einer Auktion empfehlenswert ist, stellt die Situation eines Vertrauenspartners nur einen Spezialfall dar für den viel allgemeineren Fall, der in der Praxis sehr häufig beobachtet wird: Interne Präferenzen für einzelne Bieter können dazu führen, dass nur ein oder zwei Vertragspartner für einen Auftrag zugelassen sind. Letztlich noch allgemeiner und für die hiesige Fragestellung wiederum gleichbedeutend ist die Situation, wenn schon im Markt nur ein oder zwei Anbieter für den Auftrag zu finden sind. In all diesen Fällen gilt: Mit einem einzelnen Anbieter kann keine Auktion durchgeführt werden. Der Sonderfall einer verdeckten Erstpreisauktion mit einem einzelnen Bieter sei hier als Ausnahme erwähnt und nicht weiter diskutiert. Im Fall von zwei Bietern ist eine Auktion nur dann ohne weiteres empfehlenswert, wenn die beiden Bieter nicht wissen, dass es nur zwei Bieter sind. Andernfalls muss die Wettbewerbssituation zwischen den beiden Bietern genauer analysiert werden, bevor für oder gegen die Veranstaltung einer Auktion entschieden wird.

Zusammenfassend ergeben sich die Voraussetzungen für die Veranstaltung von Auktionen wie folgt:

- Im Markt muss Wettbewerb herrschen – d.h. es darf keine Kollusion drohen und es müssen mindestens zwei Bieter vorhanden sein, wobei im Fall von zwei Bietern die beiden Bieter die Anzahl 2 der Bieter nicht kennen dürfen.
- Die Bieter müssen vergleichbar sein – sowohl bezüglich ihrer eigenen Risikoaversion als auch bezüglich der ggf. asymmetrischen Präferenzen des Kunden zwischen den Bietern

Als dritte, eher technische Voraussetzung für die Anwendung des in Teil 4 der Veröffentlichung vorgestellten „Auktionswürfels" formulieren wir noch, dass es sich um eine Auktion um ein einzelnes Verhandlungsobjekt (nicht mehrere „Lose", die von verschiedenen Bietern gewonnen werden können) handeln muss.

**Teil 3: Diskussion der Basisauktionsformen**

Die verdeckte Erstpreisauktion:

Bei der verdeckten Erstpreisauktion geben alle Bieter genau ein Angebot ab und erfahren bis zur Entscheidungsfindung durch den Kunden nichts über die Gebote der Wettbewerber. Der Auktionator sichtet alle Angebote und gibt dem Bieter mit dem besten Angebot den Zuschlag, wobei dieser seinen eigenen, angebotenen Preis erhält. Deshalb heißt die verdeckte Erstpreisauktion auch „pay-as-bid-" oder „pay-your-bid-Auktion". ([Klemperer], Seite 115).

Das Ergebnis einer verdeckten Erstpreisauktion entspricht im Erwartungswert E(FPSB) genau der im Teil 1 angeführten Betrachtung allgemeiner Erstpreisauktionen: E(FPSB) = E(X(1,n))+M.

In der Praxis hat eine verdeckte Erstpreisauktion den Beigeschmack der Informations-Exponierung: Der Auktionator erfährt die Angebotspreise aller Bieter und könnte diese auch nach der Auktion noch einmal als Verhandlungsargument nutzen, um das Ergebnis der Auktion nachzuverhandeln. Deshalb ist bei verdeckten Erstpreisauktionen besonders auf die Entscheidungsverbindlichkeit des Auktionators zu achten, die wir in der vorliegenden Veröffentlichung als uneingeschränkt vorhanden annehmen wollen.

Ein weiteres, in der Praxis beobachtetes und in der Spieltheorieliteratur diskutiertes Phänomen der verdeckten Erstpreisauktion ist der „Fluch des Gewinners" ([Holler 2020] Seite 249f). Treffen für die Bieter hohe Unsicherheiten in der Bewertung des Verhandlungsobjekts mit einer so genannten „Common Value" Situation – d.h. das Verhandlungsobjekt ist theoretisch aus Sicht aller Bieter gleich zu bewerten – zusammen, dann wird derjenige Bieter sich in einer verdeckten Erstpreisauktion durchsetzen, der sich am meisten vermeintlich zu seinen Gunsten, tatsächlich aber zu seinen Ungunsten verschätzt hat. Er erfährt mit dem Zuschlag, dass er sich am meisten verschätzt hat, und muss aber seinen eigenen, „verschätzen" Preis akzeptieren. Diese Situation gibt dem „Fluch des Gewinners" seinen Namen. Der bisweilen diskutierte Effekt, dass risikoaverse Bieter um so höhere Preise in einer verdeckten Erstpreisauktion abgeben, um den Fluch des Gewinners zu vermeiden, wird in der Praxis nur bei sehr erfahrenen Bietern beobachtet. Überwiegt die andere Risikoaversität, nämlich diejenige dagegen, den Zuschlag nicht zu erhalten, dann kommt es durchaus häufig zum Fluch des Gewinners.

Die verdeckte Zweitpreisauktion:

Bei der verdeckten Zweitpreisauktion geben alle Bieter genau ein Angebot ab und erfahren bis zur Entscheidungsfindung durch den Kunden nichts über die Gebote der Wettbewerber. Der Auktionator sichtet alle Angebote und gibt dem Bieter mit dem besten Angebot den Zuschlag, wobei er den zweitbesten Preis erhält, der angeboten wurde. Wie bereits in Teil 1 der Veröffentlichung erwähnt führt dies zur streng dominanten Strategie der Bieter, genau ihre Indifferenzpreise anzubieten (auch: „truth-revealing-machanism") ([Holler 2020] Seite 239).

Das Ergebnis einer verdeckten Zweitpreisauktion entspricht im Erwartungswert E(SPSB) genau der im Teil 1 angeführten Betrachtung allgemeiner Zweitpreisauktionen: E(SPSB) = E(X(2,n)).

In der Praxis hat eine verdeckte Zweitpreisauktion neben dem Beigeschmack der Informations-Exponierung – der noch stärker relevant ist als bei der verdeckten Erstpreisauktion denn in der verdeckten Zweitpreisauktion werden theoretisch die unverfälschten Indifferenzpreise exponiert – zusätzlich das Problem dass es in der Praxis enorme Disziplin bzw. einen hohen Ausbildungsstand und Vertrauen in die Spieltheorie aller beteiligten Entscheidungsträger erfordert, um eine verdeckte Zweitpreisauktion auch wirklich als Zweitpreisauktion konsequent durchzuführen. Die Versuchung ist hoch, am Ende doch auf den angebotenen Preis des besten Bieters zurückzugreifen und hoch muss auch das Vertrauen der Bieter sein, dass dies auf Kundenseite nicht passiert.

In der Spieltheorieliteratur wird für die verdeckten Zweitpreisauktion proklamiert, besonders anfällig für Kartellbildung zu sein (z.B. [Klemperer 2004] Abschnitt 1.9). Sprechen sich Bieter untereinander ab, können sie über die Angebotsgestaltung nicht nur genau steuern, wer den Auftrag bekommen soll und zu welchem Preis sondern darüber hinaus auch den Abweichungsanreiz der Kartellteilnehmer minimieren – indem der dezidierte Gewinner einen für die Bieter tatsächlich unattraktiven Preis bietet. Dies erfordert allerdings eine wirklich explizite Absprache unter den Bietern, die im Allgemeinen wettbewerbsrechtlich relevant ist. Liegen im Markt allerdings „nur" kollusive Tendenzen und keine expliziten Kartellabsprachen vor, dann ist nach unserer Praxiserfahrung eine verdeckte Zweitpreisauktion für die Bieter nicht dazu geeignet, eine nicht explizit abgesprochene Koordination unter den Bietern zu fördern. Im

Gegenteil: Werden seitens des Kunden die strategischen Maßnahmen gegen Kollusion – also z.B. konsequentes Bündeln aller für die Bieter in absehbarer Zeit relevanten Geschäftsvorfälle mit dem Kunden – erfolgreich angewandt um mehr als einen Bieter zum Ausbrechen aus der Kollusion anzureizen, dann befreit die verdeckte Zweitpreisauktion die abgegeben Preise komplett von jeder historisch etablierten Kollusionsmarge in den Angeboten. Es müssen allerdings mindestens zwei Bieter aus der Kollusion „ausbrechen", damit sich der Effekt im Auktionsergebnis einer verdeckten Zweitpreisauktion einstellt.

Die Englische Auktion:

Bei der Englischen Auktion bekommen die Bieter vom Auktionator Informationen über die Höhe der Gebote der anderen Bieter und verbessern ihr Gebot so lange, bis kein Bieter mehr zu einer weiteren Angebotsverbesserung bereit ist. In der Praxis industrieller Einkaufsorganisationen haben sich einige auf den ersten Blick sehr unterschiedliche Abwicklungsformen von Englischen Auktionen etabliert: Die Informationen, die den Bietern gegeben werden, können gegenüber der vollen Transparent aller - aus wettbewerbsrechtlichen Gründen immer anonymisierten – Angebotspreise auch abgeschwächte Teilinformationen sein wie zum Beispiel nur der Preis des jeweils besten Bieters oder nur der Rang, auf dem jeder Bieter selbst liegt oder auch nur eine „Ampelfarbe", wobei in jeder Farbe wiederum mehrere Ränge zusammengefasst sind. Neben diesen „Dynamischen Englischen Auktionen", bei welchen die Bieter alle selbst aktiv ihre Gebote abgeben und verbessern, hat sich auch die Form der Englischen Tickerauktion in der Praxis etabliert. Hier ruft der Auktionator aktiv in fallenden Schritten Preise ab und gibt nach jedem Schritt die Information an die Bieter zurück, ob noch mehr als ein Bieter den letzten Schritt bestätig hat oder wenn das nicht der Fall ist, dass die Auktion beendet ist. Alternativ kann der Auktionator auch die Zahl der Bieter nennen und mit jedem Schritt herunterzählen. Die Tickerform der Englischen Auktion nennt sich in der Praxis auch Japanische Auktion. Angeblich geht die Namensgebung auf den japanischen Fischmarkt zurück, wobei dort – wie am allmorgendlichen Fischmarkt in Hamburg – sicherlich eine descending clock auction als forward auction zur Anwendung kommt. Es handelt sich also am Fischmarkt um eine Holländische Auktion ([Berz 2014] Seite 40ff) und damit um eine Begriffsverwirrung. Die separate Namensgebung ist ohnehin redundant, denn bei der Englischen Tickerauktion handelt es sich – spätestens wenn man theoretisch an infinitesimal kleine Tickerschritte denkt – um die reinste Form, die Idee der Englischen Auktion umzusetzen.

Das Ergebnis einer Englischen Auktion entspricht im Erwartungswert E(Englisch) auf den ersten Blick dem einer verdeckten Zweitpreisauktion, also $E(Englisch) = E(X(2,n))$ ([Berz 2014] Seite 34). Allerdings zahlt die Transparenz der Gebotspreise darauf ein, dass die Bieter unter dem Eindruck des Wettbewerbs ihre Indifferenzpreise während der Auktion neu justieren. Dies ist vor allem dann der Fall, wenn die Kalkulation der Indifferenzpreise Unsicherheiten unterliegt und zusätzlich eine Common Value Situation gegeben ist. Sowohl der Erwartungswert als auch das Sigma der Normalverteilung, die den Indifferenzpreisen zu unterstellen ist, ändert sich also durch die Auktion. Wir bezeichnen mit Z die Ordnungsstatistiken der nach dem Preiswettbewerb unter den Bietern erhaltenen Normalverteilung und erkennen $E(Z(2,n))$ als erwartetes Ergebnis der Englischen Auktion.

Die Standardabweichung Sigma der Normalverteilung der Indifferenzpreise verringert sich in der Regel durch den Preiswettbewerb, weil die Bieter näher „zusammengetrieben" werden. Ob sich auch der Erwartungswert der Normalverteilung, und damit auch $E(Z(2,n))$ gegenüber $(E(X(2,n))$, verringert – also für den Auktionator verbessert – hängt genau von der Risikoaversion der Bieter gegenüber dem Fluch des Gewinners ab, die weiter oben im Kontext der verdeckten Erstpreisauktion bereits erwähnt wurde. Bietern, denen der Fluch des Gewinners droht weil ihre Risikoaversion gegenüber dem Verlust des Zuschlages überwiegt, hilft die Englische Auktion den Fluch des Gewinners zu vermeiden. Als Auktionator muss man sich genau im Klaren sein, ob man im Zweifel den Fluch des Gewinners vermeiden möchte oder nicht. Vermeidet man ihn, führt dieses ggf. zu einem höheren Erwartungswert $E(Z(2,n))$ als $(E(X(2,n))$. Genau aus diesem Grund haben sich in der industriellen Einkaufspraxis fast ausschließlich Formen der Englischen Auktion etabliert, bei denen die Bieter nicht wissen, wie viele andere Bieter bereits aus dem Prozess ausgestiegen sind. Damit soll $E(Z(2,n)) \leq (E(X(2,n))$ sicher gestellt werden, aber der Fluch des Gewinners wird dann nicht vermieden. Die einzige in der industriellen Einkaufspraxis gängige Auktionsform, die den Fluch des Gewinners vermeidet, ist diejenige Englische Tickerauktion, bei der die Zahl der Bieter heruntergezählt wird ([Berz 2014] Seite 98 und 103). Auch ohne der Information, wie viele andere Bieter schon ausgestiegen sind, ist übrigens die Beziehung $E(Z(2,n)) \leq (E(X(2,n))$ nicht sicher gestellt. Bleibt nämlich aufgrund des kollusiven Verhaltens der Bieter, das im Teil 2 der Veröffentlichung vorgestellt und auch im Kontext der verdeckten Zweitpreisauktion im Teil 3 diskutiert wurde, der Erwartungswert der Normalverteilung möglicherweise sogar gleich und verringert sich nur Sigma, dann wird $E(Z(2,n)) > (E(X(2,n))$ sein, weil die Indifferenzpreise „näher zusammenrücken". Dies wird in der Praxis verhindert, indem einmal abgegebene Preise nicht erhöht werden dürfen. Dann wird in einer solchen Situation allerdings gar keine Bewegung in der Auktion entstehen – ein Effekt, der durchaus auch zu beobachten ist. Man spricht von einem „Auktionsflop".

Die Hongkong Auktion:

Die Hongkong Auktion ist eine Auktion mit m Gewinnern, m > 1, die genau so abläuft wie eine Englische Tickerauktion mit m-1 Gewinnern, allerdings mit dem einen Unterschied, dass nicht nur m-1 sondern die m besten Bieter als Gewinner gekürt werden. Der letzte Bieter, der aus dem Ticker aussteigt und damit das Abbruchkriterium von nur noch m-1 verbleibenden Bietern auslöst, erfährt erst nach seinem Ausstieg davon dass er auch Gewinner ist. Deshalb kann man die Hongkong Auktion nicht (bis zum Auktionsende) „herunterzählen" sondern kann (spätestens wenn nur noch m+1 Bieter in der Auktion aktiv sind) nach jedem Tickerschritt nur die Information, ob noch mehr als m-1 Bieter vorhanden sind, an die Bieter zurückgeben.

Namensgebend für die Hongkong Auktion war eine Empfehlung des Auktionstheoretikers Klemperer für die 3G Mobilfunklizenzauktion in Hongkong ([Klemperer 2004] Abschnitt 4.4.1). Stark vereinfacht sollten m Gewinner unter n Bietern identifiziert werden. Klemperer schlug vor, nach einer verdeckten Angebotsrunde mit allen n Bietern den m Bietern mit den m besten Preisen den Zuschlag zu geben, allerdings allen zu dem aus Sicht des Auktionators im m-ten Rang stehenden Angebotspreis. Mit anderen Worten schlug Klemperer eine „Uniform Price Auction" mit dem Preis des schlechtesten Gewinners als Preis für alle Gewinner vor. Würde man den Bietern unterstellen dass sich Indifferenzpreise unter dem Eindruck von Wettbewerb nicht verändern, dann wäre das Ergebnis dieser Auktionsform exakt dasselbe wie das einer Hongkong Auktion, wie wir sie als absteigende Tickerauktion definieren.

Das Ergebnis einer Hongkong Auktion mit m Gewinnern E(Hongkong(m)) entspricht auf den ersten Blick dem einer Englischen Auktion mit m-1 Gewinnern, also m*E(Z(m,n)). Allerdings ist die Hongkong Auktion kein truth-revealing-Machanismus, da der aussteigende Bieter immer noch eine Chance hat zu gewinnen und dann den von ihm selbst angebotenen Preis ausbezahlt bekommt. Also wird, ähnlich wie in der Erstpreisauktion bzw. einer pay-as-bid-Auktion mit m Gewinnern, jeder Bieter eine Strategische Marge N > 0 auf sein Gebot aufschlagen, um nicht, wenn er „schlechtester Gewinner" wird, tatsächlich keinen Gewinn zu haben. Weiter kann man argumentieren dass N echt kleiner M ist, denn der Bieter kann unter der Voraussetzung, dass er Gewinner wird, mit einer Wahrscheinlichkeit von (m-1)/m davon ausgehen, dass nicht er sondern ein anderer Bieter den für alle Gewinner gleichen Preis („uniform-price") stellt. Das löst zwar noch nicht den durchschlagen truth-revealing-Effekt aus, gibt aber zumindest einen Anreiz, die Strategische Marge im Angebotspreis gegenüber M zu reduzieren. Mit diesem N, 0 < N < M, gilt also E(Hongkong(m)) = m*(E(Z(m,n)) + N).

Analog zu Teil 1 der Veröffentlichung können wir uns jetzt einer Grenze H für die Größe der Strategischen Marge N annähern bis zu der eine Hongkong Auktion mit m Gewinnern besser ist als eine Englische Auktion mit m Gewinnern. In Form der Erwartungswerte der betreffenden Ordnungsstatistiken gilt: N < H := E(Z(m+1,n)) – E(Z(m,n)).

Nimmt man – wie in Teil 1 der Veröffentlichung - für eine Näherung H' von H näherungsweise die Gleichverteilung auf (-sqrt(3) *Sigma, sqrt(3) *Sigma) als Wahrscheinlichkeitsverteilung der Indifferenzpreise an, so können wir wieder mit den als Betafunktionen erkannten Ordnungsstatistiken Y' der Gleichverteilung rechnen anstelle von Z. Analog zum Teil 1 der Veröffentlichung kann man leicht herleiten, dass für alle i mit 0 < i < n+1 gilt: E(Y'(i,n)) = 2*sqrt(3)*Sigma* (2*i-1-n)/(2*n+2) und man erhält H' = E(Y'(m+1,n)) – E(Y'(m,n)) = 2*sqrt(3)*Sigma* (2*(m+1)-1-n)/(2*n+2) - 2*sqrt(3)*Sigma* (2*m-1-n)/(2*n+2) = 2*sqrt(3)*Sigma/(n+1) = G'.

Für m > 1 ist nicht nur der Vergleich zwischen Hongkong und Englischer Auktion interessant, sondern auch der mit einer pay-as-bid-Auktion mit m Gewinnern. Diese hat im Erwartungswert das Ergebnis (Summe aller E(X(i,n)) für i = 1, ..., m) + m*M. Für m = 2 also E(X(1,n))+E(X(1,n))+2*M. Leider kennt die Auktionstheorie für m > 1 kein Erlösäquivalenztheorem wie für m = 1, so dass wir nicht wie in Teil 1 der Veröffentlichung einen Rückschluss auf M oder N risikoneutraler Bieter ziehen können. Trotzdem ist es sehr Aufschlussreich für die Abwägung, ob man als Auktionator eine Hongkong Auktion wählen soll, sich die Erwartungswerte für m = 2 unter der Gleichverteilungsannahme einmal näher anzuschauen:

E(Englisch(m=2)) = 2* E(Y'(3,n)) = 4*sqrt(3)*Sigma*(5-n)/(2n+2)
E(Hongkong(m=2)) = 2*(E(Y'(2,n)) + N) =  4*sqrt(3)*Sigma*(3-n)/(2n+2)+2*N

E(pay-as-bid(m=2)) = E(Y'(1,n))+ E(Y'(2,n))+2*M = 2*sqrt(3)*Sigma*((1-n)/(2n+2)+ (3-n)/(2n+2))+2*M = 4*sqrt(3)*Sigma*(2-n)/(2n+2)+2*M

Beispielsweise ergibt sich für n = 6 und Sigma = 1/(2*sqrt(3)):
E(Englisch) = -1/7
E(Hongkong) = -3/7+2*N
E(pay-as-bid) = -4/7+2*M

In der Abwägung zwischen Hongkong und Englischer Auktion muss N < 1/7 = G' sein, um für die Hongkong Auktion zu sprechen.

In der Abwägung zwischen pay-as-bid und Englischer Auktion muss M < 1/14 sein, um für die pay-as-bid Auktion zu sprechen.

In der Abwägung zwischen Hongkong und pay as bid Auktion schließlich muss N < M - 1/14 sein, um für die Hongkong Auktion zu sprechen.

Um sich diese Zusammenhänge illustrativ vor Augen zu führen, betrachten wir die durchschnittlich pro Gewinner erzielten Auktionsergebnisse auf dem Intervall (-1/2, 1/2):

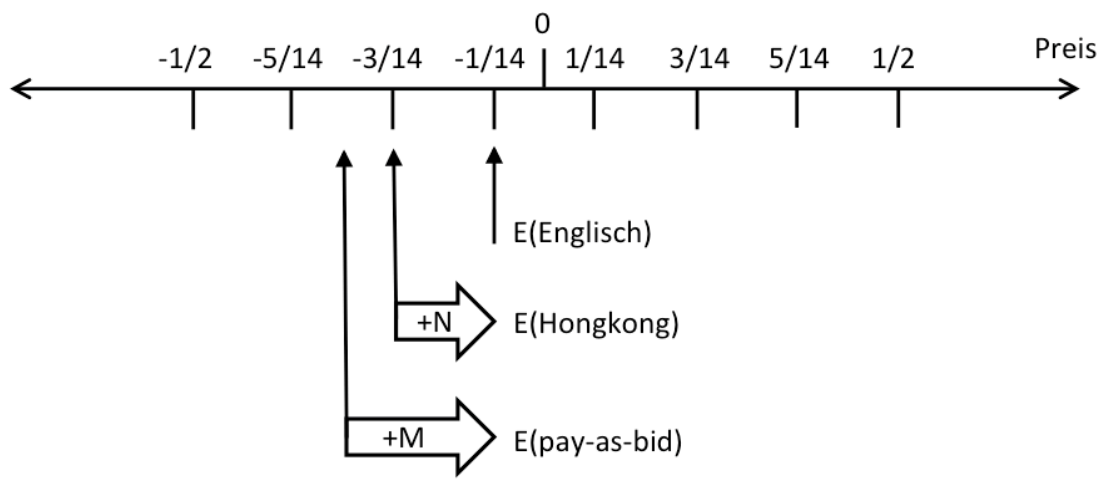

Mit Hilfe der bereits erwähnten Formel E(Y'(i,n)) = 2*sqrt(3)*Sigma* (2*i-1-n)/(2*n+2) lassen sich diese Betrachtungen bei Gleichverteilungsannahme leicht auch auf m > 2 anwenden. Wir wollen uns allerdings der wohl praxisnäheren Betrachtung von normalverteilten Indifferenzpreisen zuwenden und auf m = 2 fokussieren, da wir die Hongkong Auktion im Teil 4 der Veröffentlichung ausschließlich als Ausschlussrunde mit zwei Gewinnern verwenden, die beide an einer finalen Auktionsrunde teilnehmen sollen.

Bei der Berechnung der exakten Werte für H = E(Z(3,n)) – E(Z(2,n)) mit dem Computeralgebrasystem Maple fällt uns wie schon im Teil 1 der Veröffentlichung – dort für die 1. und 2. Ordnungsstatistik - auch für die 3. Ordnungsstatistik einer Normalverteilung mit Standardabweichung Sigma auf, dass der Erwartungswert proportional zu Sigma ist. Auch für die 3. Ordnungsstatistik folgt dies direkt aus der Skalierungseigenschaft der Normalverteilung, siehe Teil 1 der Veröffentlichung. Die Faktoren h(n) zur Berechnung von H = h(n)*Sigma für n = 3, ..., 100 können dem Anhang entnommen werden.

<u>Die Holländische Auktion:</u>

Bei der Holländischen Auktion ruft der Auktionator aktiv in steigenden Schritten Preise ab und gibt nach jedem Schritt die Information an die Bieter zurück, ob ein Bieter den letzten Schritt bestätig hat und wenn das der Fall ist, dass die Auktion beendet ist. Analog zur Englischen Tickerauktion ist es die theoretisch reinste Form, die Idee der Holländischen Auktion umzusetzen, infinitesimal kleine Tickerschritte vorzunehmen. Der Startpreis einer Holländischen Auktion liegt theoretisch beliebig niedrig.

Das Ergebnis einer Holländischen Auktion entspricht im Erwartungswert E(Dutch) in erster Näherung dem einer Erstpreisauktionen: E(Dutch) = E(FPSB) = E(X(1,n))+M. Allerdings zahlen zwei Effekte auf eine Reduzierung von M ein, die in der Praxis eine große Rolle spielen. Beide Effekte sind umso größer, je höher die Risikoaversion der Bieter dagegen, den Zuschlag nicht zu erhalten, ist.

Zum einen fällt bei der Holländischen Auktion die im Kontext der verdeckten Erstpreisauktion diskutierte Informations-Exponierung für den Bieter weg. Der Bieter weiß in dem Moment seiner Bestätigung eines Tickerschrittes, dass er damit das Geschäft sicher abschließt. Damit kann er, wenn er die Signalwirkung seines Angebotspreises an den Kunden reflektiert, den Fall das Geschäft gar nicht zu erhalten ausschließen. Man kann diese Eigenschaft der Holländischen Auktion auch als „Duale Transparenz" verstehen. Während in der Englischen Auktion, deren Idee von der Preistransparenz lebt, die Bieter immer wissen dass es noch Wettbewerber gibt die den Preis „mitgehen", ist es in der Holländischen Auktion umgekehrt: Solange die Tickerschritte der Holländischen Auktion nicht bestätigt werden wissen alle beteiligten Bieter, dass KEIN Wettbewerber diese Preise „mitgeht".

Der zweite Grund, warum Bieter in einer Holländischen Auktion mit niedrigen Strategischen Margen M anbieten als in der verdeckten Erstpreisauktion ist, dass die Entscheidungsverbindlichkeit des Auktionators bei der Holländischen Auktion außer Frage steht. Der Auktionator erhält bei der Holländischen Auktion nur von einem einzelnen Bieter überhaupt einen Preis. Die Holländische Auktion generiert kein Verhandlungsargument für den Auktionator diesem Gewinner der Auktion gegenüber, mit dem er eine Nachverhandlung eröffnen könnte. Das wissen auch alle anderen Bieter und können daraus schließen, dass der Gewinner der Holländischen Auktion das Geschäft ultimativ erhält und keine Nachverhandlungen mehr stattfinden können. Also können die Bieter auf Risikomargen oder „Puffer", die sie in ihren Angebotspreisen bzw. in der Strategischen Marge M für eventuelle Nachverhandlungen einpreisen würden, vollständig verzichten.

**Teil 4: Der Auktionswürfel**

Im vierten Teil der Veröffentlichung setzen wir die Erfüllung der im Teil 2 formulierten Kriterien für die Veranstaltung einer Auktion voraus und nutzen die in Teil 3 vorgestellten Basisauktionsformen, um in Abhängigkeit der drei Größen n (Anzahl der Bieter), M (Strategische Marge der Bieter in einer Erstpreisauktion, reziprok zu Risikoaversion den Zuschlag nicht zu erhalten) und Sigma (Standardabweichung der für die Indifferenzpreise der Bieter identisch – d.h. insbesondere mit gleichem Erwartungswert – angenommenen Normalverteilung) jeweils ein Auktionsdesign als das im Erwartungswert für den Auktionator erfolgsprechendste zu identifizieren.

Die drei Größen n, M und Sigma spannen dabei den dreidimensionalen Würfel auf, wobei wir uns für die pragmatischere Anwendbarkeit in der Praxis auf jeweils zwei Ausprägungen in jeder der drei Dimensionen einlassen: „klein" und „groß". Dies soll jeweils so zu verstehen sein, dass die in Teil 1 gefundene Ungleichungs-Bedingungen zur Abwägung zwischen Erst- und Zweitpreis Auktionen entweder klar beantwortet oder als indifferent zu betrachten ist. Bezüglich n genügt „groß" als „n > 4" zu verstehen, um sinnvollerweise eine Englische bzw. Hongkong Auktion anzuwenden. Ob hierbei der Fluch des Gewinners vermieden werden soll, also die Zahl der verbleibenden Bieter heruntergezählt wird, wollen wir mit dem Auktionswürfel nicht weiter präjustieren und kann in den betreffenden Ecken des Würfels weiterhin zusätzlich entschieden werden.

Ausgangpunkt aller Überlegungen für die empfohlenen Auktionsdesigns in den „Ecken" des Auktionswürfels ist die so genannte „Klemperer Auktion", die der Auktionstheoretiker Klemperer selbst „Anglo – Dutch Auction" nennt ([Klemperer 2004], Seite 116). Wenn keine weiteren Informationen über die Wettbewerbssituation der Bieter, also deren Indifferenzpreise und deren Risikoaversion, bekannt sind, dann sei nach Klemperer diese Auktionsform besonders „robust" und im Zweifel zu empfehlen. Letztlich besteht die Idee darin, in einer ersten „englischen" Runde per Preistransparenz Einfluss auf die Indifferenzpreise zu nehmen, so dass in der Nomenklatur des 2. Teils der vorliegenden Veröffentlichung mit den Ordnungsstatistiken Z anstelle von X gerechnet werden kann. Mit den verbleibenden Bietern wird dann eine Holländische Auktion zur Ermittlung des finalen Gewinners durchgeführt, um den niedrigsten Indifferenzpreis zu nutzen - unter Inkaufnahme einer Strategischen Marge M.

In der Praxis des industriellen Einkaufs hat sich schon seit vielen Jahren eine Adaption der Klemperer Auktion etabliert, nämlich eine Hongkong Auktion mit zwei Gewinnern, die in einer daran anschließenden Holländischen Auktion den finalen Gewinner unter sich ausmachen. Dieses Design „Hongkong – Dutch" wird der Erfahrung gerecht, dass man oft nur eine Hand voll Bieter zu Verfügung hat und dabei meistens schon der Dritte Bieter nur einen signifikant höheren Preis bieten kann als der beste und auch der zweitbeste Bieter. In der Nomenklatur und gemäß der Überlegungen von Teil 3 der Veröffentlichung heißt das, in der Praxis wird den Bietern meist eine geringere Strategische Marge N als die Grenze H unterstellt, unter der eine Hongkong Auktion ein für den Auktionator besseres Ergebnis liefert als die Englische Auktion mit zwei Gewinnern.

In zwei der acht „Ecken" des Auktionswürfels wird genau dieses „Hongkong - Dutch" Auktionsdesign empfohlen. In den anderen sechs Ecken kommt es aufgrund der besonderen Konstellation der jeweiligen Ecke zu weiteren Adaptionen. Abschließend wollen wir die acht Ecken des Würfels im Einzelnen diskutieren:

Konstellation n > 3, Risikoaversion der Bieter gering und Sigma klein:

Kommen geringe Risikoaversion der Bieter dagegen den Zuschlag nicht zu erhalten, genügend viele Bieter und nahe beieinander liegende Preisniveaus zusammen, sprechen alle auktionstheoretischen Aspekte die in den Teilen 1, 2 und 3 der Veröffentlichung ausführlich diskutiert wurden für ein Englische Auktion, vorzugsweise als Ticker der bis zum Ende mit einem Gewinner läuft.

→ Empfohlenes Auktionsdesign: **Englische Auktion**

Konstellation n > 3, Risikoaversion der Bieter gering und Sigma groß:

Je weiter dabei allerdings die Preisniveaus als weit gespreizt zu erwarten sind, desto höher liegt dabei die in Teil 1 der Veröffentlichung diskutierte Grenze G = g(n)*Sigma für M, unter der eine Erstpreisauktion als finale Preisfindung das bessere Ergebnis im Erwartungswert liefert. Ganz nach Klemperer bzw. den Eingangs in Teil 4 beschriebenen Erfahrungen aus der Praxis empfehlen wir in dieser Situation eine Hongkong Auktion, die zwei Teilnehmer der finalen Entscheidungsrunde qualifizieren. Diese Entscheidungsrunde sollte als Erstpreisauktion den Preis des besseren der beiden Bieter nutzen, unter Inkaufnahme seiner Strategischen Marge M. Die Abwägung, ob diese zweite Runde als Holländische Auktion oder als verdeckte Erstpreisauktion stattfinden soll liegt auktionstheoretisch nahe beieinander. Wie in Teil 3 der Veröffentlichung für die Holländische Auktion diskutiert, erwarten wir nur bei ausreichender Risikoaversion der Bieter eine reduzierte Strategische Marge in der Holländischen Auktion. Umgekehrt führt bei geringer Risikoaversion die Ankündigung eine Holländischen Auktion bei manchen Bietern schon zu Abwehrreaktionen und ggf. sogar der Nichtteilnahme an der Vergabe. Deshalb empfehlen wir hier eine verdeckte Erstpreisauktion.

→ Empfohlenes Auktionsdesign: **Hongkong - FPSB**

Konstellation n > 3, Risikoaversion der Bieter hoch und Sigma groß:

Bei hoher Risikoaversion der Bieter dagegen den Zuschlag nicht zu erhalten ist mit kleinen strategischen Margen zu rechnen und als finale, zweite Runde in jedem Fall eine Erstpreisauktion empfehlenswert. Sind genügend viele Bieter vorhanden, kann die erste Runde als Hongkong Ticker stattfinden, um die Indifferenzpreise gemäß den Überlegungen von Teil 3 der Veröffentlichung zu beeinflussen. Die finale

Entscheidungsrunde kann bei ausreichender Risikoaversion der Bieter als Holländische Auktion stattfinden, um die Strategische Marge zu reduzieren.

→ Empfohlenes Auktionsdesign: **Hongkong - Dutch**

Konstellation n > 3, Risikoaversion der Bieter hoch und Sigma klein:

Die Spreizung der Preisniveaus spielt im zuvor diskutierten Fall keine Rolle. Selbst wenn bei kleinem M und kleinem Sigma die Ungleichung M < g(n)*Sigma in Frage steht, sind beide Seiten der Ungleichung klein. D.h., der Nachteil, der mit einer Erstpreisauktion realisiert wird, bleibt überschaubar.

→ Empfohlenes Auktionsdesign: **Hongkong - Dutch**

Konstellation n= 2 oder 3, Risikoaversion der Bieter gering und Sigma klein:

Wenn zu wenige Bieter für einen Hongkong Ticker vorhanden sind und hohe Risikoaversion zudem zu Ablehnung gegen eine Holländische Auktion führen sind zwei Runden verdeckter Erstpreisauktionen zu empfehlen. Die erste der beiden Runde ist im engeren Sinn gar keine „Erstpreisauktion" sondern eine „pay-as-bid-Auktion" mit zwei Gewinnern und wird hier nur zur Vereinfachung der Bezeichnungen auch als verdeckte Erstpreisauktion (FPSB) bezeichnet. Beide Runden sind dabei als verdeckte Auktionen mit Posttransparenz durchzuführen, d.h. den Bietern wird nach der Entscheidung der Preis der anderen Bieter offen gelegt. Dies dient der in Teil 3 angesprochenen Entscheidungsverbindlichkeit, die den Bietern gegenüber schon vor der Auktion glaubwürdig gemacht werden muss. Für die Ausschlussrunde mit zwei Gewinnern, bei der ggf. nur zwei Bieter beteiligt sind, ist den Bietern von vorne herein nur anzukündigen dass nur die beiden Gewinner jeweils den Preis des anderen Gewinners erfahren.

→ Empfohlenes Auktionsdesign: **FPSB - FPSB**

Konstellation n= 2 oder 3, Risikoaversion der Bieter gering und Sigma groß:

Ist als einzige Abweichung zum zuvor diskutierten Fall mit einer weiten Spreizung zwischen den Preisniveaus zu rechnen sollte die Posttransparenz der ersten Runde nur in der Ranginformation der beiden Gewinner bestehen, um den besseren Bieter nicht unnötig in Sicherheit zu wiegen. Wir kürzen diese Version der pay-as-bid-Auktion mit zwei Gewinnern mit „FPSB\R" ab.

→ Empfohlenes Auktionsdesign: **FPSB\R - FPSB**

Konstellation n= 2 oder 3, Risikoaversion der Bieter hoch und Sigma klein:

Bei hoher Risikoaversion der Bieter ist mit kleinen strategischen Margen zu rechnen und als finale, zweite Runde in jedem Fall eine Erstpreisauktion zu empfehlen. Diese liefert, wiederum aufgrund der hohen Risikoaversion der Bieter, als Holländische eine reduzierte Strategische Marge M. Sind nicht genügend viele Bieter für eine Hongkong Auktion vorhanden, muss die Qualifikationsrunde als verdeckte Erstpreisauktion gespielt werden.

→ Empfohlenes Auktionsdesign: **FPSB - Dutch**

Konstellation n= 2 oder 3, Risikoaversion der Bieter hoch und Sigma groß:

Die Spreizung der Preisniveaus spielt im zuvor diskutierten Fall für die Abwägung Erst- oder Zweitpreisauktion der finalen Runde keine Rolle. Selbst wenn bei kleinem M und kleinem Sigma die Ungleichung M < g(n)*Sigma in Frage steht, sind beide Seiten der Ungleichung klein. D.h., der Nachteil, der mit einer Erstpreisauktion realisiert wird, bleibt überschaubar.

Die Posttransparenz der Ausschlussrunde ist auch bei großer, zu erwartender Spreizung zwischen den Preisniveaus mit transparenter Preisinformation der beiden Gewinner nicht für das Auktionsergebnis schädlich. Aufgrund der hohen Risikoaversion wird ein großer Preisabstand den zurückliegenden Bieter zusätzlich anspornen. Der führende Bieter erfährt zwar einen großen Abstand, weiß aber gleichzeitig, dass sein Kontrahent seinen Preis jetzt kennt und muss sich deshalb auch nochmal bewegen.

→ Empfohlenes Auktionsdesign: **FPSB - Dutch**

# Auction Cube ®

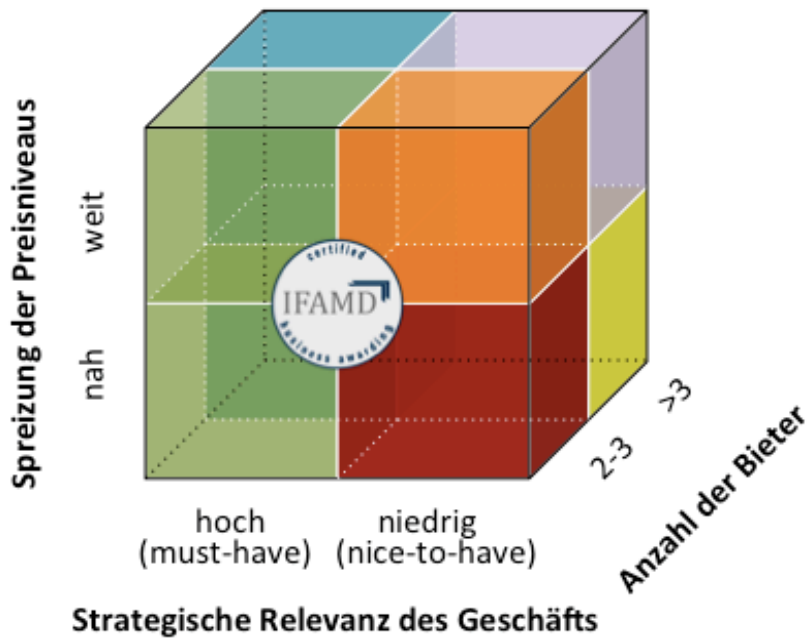

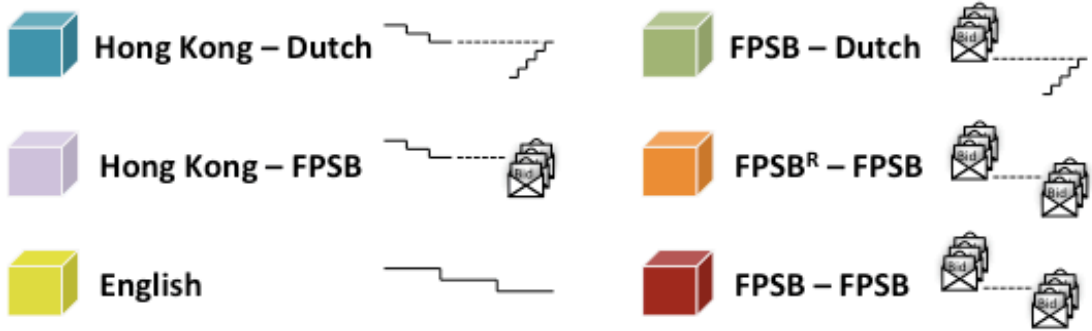

**Resümee und Ausblick:**

Für Auktionen mit einem Los und unter Bietern, die vergleichbar sind und sich gegenseitig uneingeschränkt in Wettbewerb begeben, liefert der Auktionswürfel eine eindeutige Empfehlung zum anzuwendenden Auktionsdesign. Ist mindestens eines der drei Kriterien, die gegen die Veranstaltung einer Auktion sprechen, erfüllt, dann sind andere Entscheidungs- und Verhandlungsmechanismen vorzuziehen wie zum Beispiel take-it-or-leave-it-Ketten, take-it-or-auction-Angebote, first-call-Optionen, last-call-Optionen oder Angebotsrunden mit verdeckten Zielpreisen die zu einem direkten Zuschlag führen um nur die wichtigsten der Basisverhandlungsformen zu nennen, die dann in Frage kommen. Dies ist der Fall wenn interne Präferenzen für einzelne Bieter so hoch sind, dass man nicht mehr von vergleichbaren Bietern sprechen kann oder wenn kollusives Verhalten der Bieter mit dem Verhandlungsdesign taktisch zu adressieren ist. Wenn mehr als ein Los zur Disposition stehen, öffnet sich das ganze Feld der kombinatorischen Ausschreibungen. Hier kann der Auktionswürfel eine Orientierung geben bezüglich der Entscheidungsverfahren in den einzelnen Losen. Ein zusätzliches Metadesign muss dann aber noch die Synchronisation der Entscheidungsfindung der einzelnen Lose und zugelassener Bündel von Losen definieren.

**Notation:**

| | |
|---|---|
| NV(e,s) | Normalverteilung mit Erwartungswert e und Standardabweichung s |
| n | Anzahl der Bieter |
| E(.) | Erwartungswert einer Wahrscheinlichkeitsverteilung bzw. des Ergebnisses einer Auktionsform |
| X | Ordnungsstatistiken einer Normalverteilung |
| X(k,n) | k-te Ordnungsstatistik einer Normalverteilung mit Stichprobenlänge n |
| M | Strategische Marge in einer Erstpreisauktion |
| G | Grenze für M zur Abwägung zwischen Erst- und Zweitpreisauktion unter Normalverteilungsannahme |
| Y | Ordnungsstatistiken der Gleichverteilung auf (0, 1) |
| Y' | Ordnungsstatistiken der Gleichverteilung auf (-1/2, 1/2) |
| Y'' | Ordnungsstatistiken der Gleichverteilung auf (-sqrt(3) *Sigma, sqrt(3) *Sigma)) |
| G' | Grenze für M zur Abwägung zwischen Erst- und Zweitpreisauktion unter Gleichverteilungsannahme, bei beliebig vorgegebenem Sigma |
| g(n) | Faktoren zur Berechnung von G in Abhängigkeit von n und Sigma |
| Z | Ordnungsstatistiken einer Normalverteilung von Indifferenzpreisen, die unter dem Eindruck von Wettbewerbspreisen justiert wurden |
| m | Anzahl der Gewinner in einer Auktion |
| N | Strategische Marge in einer Hongkong Auktion |
| H | Grenze für N zur Abwägung zwischen Hongkong und Englischer Auktion unter Normalverteilungsannahme |
| H' | Grenze für N zur Abwägung zwischen Hongkong und Englischer Auktion unter Gleichverteilungsannahme, bei beliebig vorgegebenem Sigma |
| h(n) | Faktoren zur Berechnung von H in Abhängigkeit von n und Sigma |

**Literatur**

[Milgrom 2004] Milgrom, Paul; Putting Auction Theory to Work; Cambridge University Press; 2004

[Klemperer 2004] Klemperer, Paul; Auctions: Theory and Practice; Princeton University Press; 2004

[Berz 2014] Berz, Gregor; Spieltheoretische Verhandlungs- und Auktionsstrategien; Schäffer Poeschel; 2007, 2. Auflage 2014

[Kabluchko 2017] Kabluchko, Zakhar; Skript zur Vorlesung „Mathematische Statistik" in der Version vom 22.02.2017; Universität Münster, Institut für Mathematische Stochastik; 2017. Link: https://www.uni-muenster.de/Stochastik/kabluchko/Skripte/Skript_Math_Statistik_Version_22_02_2017.pdf

[Holler 2020] Holler, Manfred J., Klose-Ullmann, Barbara; Scissors and Rock – Game Theory for Those Who Manage; Springer; 2020

**Anhang**

Formel für die Dichtefunktion der i. Ordnungsstatistik mit Stichprobenlänge n einer stetigen Zufallsvariable ([Kabluchko 2017] Satz 1.6.1):

**Satz 1.6.1.** Seien $X_1, X_2, \ldots, X_n$ unabhängige und identisch verteilte Zufallsvariablen, die absolut stetig sind mit Dichte $f$ und Verteilungsfunktion $F$. Es seien

$$X_{(1)} \leq X_{(2)} \leq \ldots \leq X_{(n)}$$

die Ordnungsstatistiken. Dann ist die Dichte der Zufallsvariable $X_{(i)}$ gegeben durch

$$f_{X_{(i)}}(t) = \frac{n!}{(i-1)!(n-i)!} f(t) F(t)^{i-1} (1-F(t))^{n-i}.$$

Die Tabellen auf den folgenden Seiten sind ausschließlich mit Maple gerechnete Werte für die Normalverteilung NV(0, Sigma).

| Sigma | n | G' = H' = = 2*sqrt(3)* *Sigma/(n+1) | E(X(1,n)) | E(X(2,n)) | E(X(3,n)) | G = = E(X(2,n)) - E(X(1,n)) | (G'-G)/G | H = = E(X(3,n)) - E(X(2,n)) | H'/H |
|---|---|---|---|---|---|---|---|---|---|
| 0,1 | 2 | 0,115470054 | -0,056418958 | 0,056418958 | n.a. | 0,112837917 | 2,3326708759% | n.a. | n.a. |
| 0,2 | 2 | 0,230940108 | -0,112837917 | 0,112837917 | n.a. | 0,225675833 | 2,3326708033% | n.a. | n.a. |
| 0,3 | 2 | 0,346410162 | -0,169256875 | 0,169256875 | n.a. | 0,33851375 | 2,3326708335% | n.a. | n.a. |
| 0,4 | 2 | 0,461880215 | -0,225675833 | 0,225675833 | n.a. | 0,451351667 | 2,3326708033% | n.a. | n.a. |
| 0,5 | 2 | 0,577350269 | -0,282094792 | 0,282094792 | n.a. | 0,564189584 | 2,3326707489% | n.a. | n.a. |
| 0,6 | 2 | 0,692820323 | -0,33851375 | 0,33851375 | n.a. | 0,6770275 | 2,3326708335% | n.a. | n.a. |
| 0,7 | 2 | 0,808290377 | -0,394932708 | 0,394932708 | n.a. | 0,789865416 | 2,3326708681% | n.a. | n.a. |
| 0,8 | 2 | 0,923760431 | -0,451351667 | 0,451351667 | n.a. | 0,902703334 | 2,3326708033% | n.a. | n.a. |
| 0,9 | 2 | 1,039230485 | -0,507770626 | 0,507770626 | n.a. | 1,015541251 | 2,3326707328% | n.a. | n.a. |
| 1 | 2 | 1,154700538 | -0,564189584 | 0,564189584 | n.a. | 1,128379167 | 2,3326708033% | n.a. | n.a. |
| 2 | 2 | 2,309401077 | -1,128379167 | 1,128379167 | n.a. | 2,256758334 | 2,3326708033% | n.a. | n.a. |
| 3 | 2 | 3,464101615 | -1,69256875 | 1,69256875 | n.a. | 3,3851375 | 2,3326708335% | n.a. | n.a. |
| 4 | 2 | 4,618802154 | -2,256758334 | 2,256758334 | n.a. | 4,513516668 | 2,3326708033% | n.a. | n.a. |
| 5 | 2 | 5,773502692 | -2,820947918 | 2,820947918 | n.a. | 5,641895836 | 2,3326707852% | n.a. | n.a. |
| 6 | 2 | 6,92820323 | -3,385137501 | 3,385137501 | n.a. | 6,770275002 | 2,3326708033% | n.a. | n.a. |
| 7 | 2 | 8,082903769 | -3,949327084 | 3,949327084 | n.a. | 7,898654168 | 2,3326708163% | n.a. | n.a. |
| 8 | 2 | 9,237604307 | -4,513516668 | 4,513516668 | n.a. | 9,027033336 | 2,3326708033% | n.a. | n.a. |
| 9 | 2 | 10,39230485 | -5,077706252 | 5,077706252 | n.a. | 10,1554125 | 2,3326707932% | n.a. | n.a. |
| 10 | 2 | 11,54700538 | -5,641895835 | 5,641895835 | n.a. | 11,28379167 | 2,3326708033% | n.a. | n.a. |
| 20 | 2 | 23,09401077 | -11,28379167 | 11,28379167 | n.a. | 22,56758334 | 2,3326708033% | n.a. | n.a. |
| 30 | 2 | 34,64101615 | -16,9256875 | 16,9256875 | n.a. | 33,851375 | 2,3326708335% | n.a. | n.a. |
| 40 | 2 | 46,18802154 | -22,56758334 | 22,56758334 | n.a. | 45,13516668 | 2,3326708033% | n.a. | n.a. |
| 50 | 2 | 57,73502692 | -28,20947918 | 28,20947918 | n.a. | 56,41895836 | 2,3326707852% | n.a. | n.a. |
| 60 | 2 | 69,2820323 | -33,85137501 | 33,85137501 | n.a. | 67,70275002 | 2,3326708033% | n.a. | n.a. |
| 70 | 2 | 80,82903769 | -39,49327084 | 39,49327084 | n.a. | 78,98654168 | 2,3326708163% | n.a. | n.a. |
| 80 | 2 | 92,37604307 | -45,13516668 | 45,13516668 | n.a. | 90,27033336 | 2,3326708033% | n.a. | n.a. |

| Sigma | n | G' = H' = = 2*sqrt(3)* *Sigma/(n+1) | E(X(1,n)) | E(X(2,n)) | E(X(3,n)) | G = = E(X(2,n)) - - E(X(1,n)) | (G'-G)/G | H = = E(X(3,n)) - - E(X(2,n)) | (H'-H)/H |
|---|---|---|---|---|---|---|---|---|---|
| 90 | 2 | 103,9230485 | -50,77706252 | 50,77706252 | n.a. | 101,554125 | 2,3326707932% | n.a. | n.a. |
| 100 | 2 | 115,4700538 | -56,41895835 | 56,41895835 | n.a. | 112,8379167 | 2,3326708033% | n.a. | n.a. |

| Sigma | n | G' = H' = = 2*sqrt(3)* *Sigma/(n+1) | E(X(1,n)) | E(X(2,n)) | E(X(3,n)) | G = = E(X(2,n)) - - E(X(1,n)) | (G'-G)/G | H = = E(X(3,n)) - - E(X(2,n)) | (H'-H)/H |
|---|---|---|---|---|---|---|---|---|---|
| 0,1 | 3 | 0,08660254 | -0,084628438 | 0 | 0,084628437 | 0,084628438 | 2,3326708094% | 0,084628437 | 2,33326709303% |
| 0,2 | 3 | 0,173205081 | -0,169256875 | 0 | 0,169256875 | 0,169256875 | 2,3326707731% | 0,169256875 | 2,3326707731% |
| 0,3 | 3 | 0,259807621 | -0,253885313 | 0 | 0,253885313 | 0,253885313 | 2,3326707932% | 0,253885313 | 2,3326708335% |
| 0,4 | 3 | 0,346410162 | -0,33851375 | 0 | 0,33851375 | 0,33851375 | 2,3326708033% | 0,33851375 | 2,3326708335% |
| 0,5 | 3 | 0,433012702 | -0,423142188 | 0 | 0,423142188 | 0,423142188 | 2,3326707852% | 0,423142188 | 2,3326707126% |
| 0,6 | 3 | 0,519615242 | -0,507770625 | 0 | 0,507770625 | 0,507770625 | 2,3326708134% | 0,507770625 | 2,3326708134% |
| 0,7 | 3 | 0,606217783 | -0,592399062 | 0 | 0,592399062 | 0,592399062 | 2,3326708854% | 0,592399062 | 2,3326709199% |
| 0,8 | 3 | 0,692820323 | -0,6770275 | 0 | 0,6770275 | 0,6770275 | 2,3326708335% | 0,6770275 | 2,3326708335% |
| 0,9 | 3 | 0,779422863 | -0,761655938 | 0 | 0,761655939 | 0,761655938 | 2,3326707261% | 0,761655939 | 2,3326706454% |
| 1 | 3 | 0,866025404 | -0,846284375 | 0 | 0,846284375 | 0,846284375 | 2,3326708094% | 0,846284375 | 2,3326708094% |
| 2 | 3 | 1,732050808 | -1,69256875 | 0 | 1,69256875 | 1,69256875 | 2,3326708335% | 1,69256875 | 2,3326708335% |
| 3 | 3 | 2,598076211 | -2,538853126 | 0 | 2,538853126 | 2,538853126 | 2,3326707932% | 2,538853126 | 2,3326707932% |
| 4 | 3 | 3,464101615 | -3,385137501 | 0 | 3,385137501 | 3,385137501 | 2,3326708033% | 3,385137501 | 2,3326708033% |
| 5 | 3 | 4,330127019 | -4,231421876 | 0 | 4,231421876 | 4,231421876 | 2,3326708094% | 4,231421876 | 2,3326708094% |
| 6 | 3 | 5,196152423 | -5,077706252 | 0 | 5,077706252 | 5,077706252 | 2,3326707932% | 5,077706252 | 2,3326707932% |
| 7 | 3 | 6,062177826 | -5,923990627 | 0 | 5,923990627 | 5,923990627 | 2,3326707990% | 5,923990627 | 2,3326707990% |
| 8 | 3 | 6,92820323 | -6,770275002 | 0 | 6,770275002 | 6,770275002 | 2,3326708033% | 6,770275002 | 2,3326708033% |
| 9 | 3 | 7,794228634 | -7,616559377 | 0 | 7,616559377 | 7,616559377 | 2,3326708067% | 7,616559377 | 2,3326708067% |
| 10 | 3 | 8,660254038 | -8,462843752 | 0 | 8,462843752 | 8,462843752 | 2,3326708094% | 8,462843752 | 2,3326708094% |
| 20 | 3 | 17,32050808 | -16,9256875 | 0 | 16,9256875 | 16,9256875 | 2,3326708335% | 16,9256875 | 2,3326708335% |
| 30 | 3 | 25,98076211 | -25,38853126 | 0 | 25,38853126 | 25,38853126 | 2,3326707932% | 25,38853126 | 2,3326707932% |
| 40 | 3 | 34,64101615 | -33,85137501 | 0 | 33,85137501 | 33,85137501 | 2,3326708033% | 33,85137501 | 2,3326708033% |
| 50 | 3 | 43,30127019 | -42,31421876 | 0 | 42,31421876 | 42,31421876 | 2,3326708094% | 42,31421876 | 2,3326708094% |
| 60 | 3 | 51,96152423 | -50,77706252 | 0 | 50,77706252 | 50,77706252 | 2,3326707932% | 50,77706252 | 2,3326707932% |
| 70 | 3 | 60,62177826 | -59,23990627 | 0 | 59,23990627 | 59,23990627 | 2,3326707990% | 59,23990627 | 2,3326707990% |
| 80 | 3 | 69,2820323 | -67,70275002 | 0 | 67,70275002 | 67,70275002 | 2,3326708033% | 67,70275002 | 2,3326708033% |
| 90 | 3 | 77,94228634 | -76,16559377 | 0 | 76,16559377 | 76,16559377 | 2,3326708067% | 76,16559377 | 2,3326708067% |
| 100 | 3 | 86,60254038 | -84,62843752 | 0 | 84,62843752 | 84,62843752 | 2,3326708094% | 84,62843752 | 2,3326708094% |

| Sigma | n | G' = H' = = 2*sqrt(3)* *Sigma/(n+1) | E(X(1,n)) | E(X(2,n)) | E(X(3,n)) | G = = E(X(2,n)) - - E(X(1,n)) | (G'-G)/G | H = = E(X(3,n)) - - E(X(2,n)) | (H'-H)/H |
|---|---|---|---|---|---|---|---|---|---|
| 0,1 | 4 | 0,069282032 | -0,102937537 | -0,029701138 | 0,029701138 | 0,073236399 | -5,3994555634% | 0,059402277 | 16,6319480300% |
| 0,2 | 4 | 0,138564065 | -0,205875075 | -0,059402276 | 0,059402276 | 0,146472798 | -5,3994554730% | 0,118804553 | 16,6319482852% |
| 0,3 | 4 | 0,207846097 | -0,308812612 | -0,089103415 | 0,089103415 | 0,219709197 | -5,3994555246% | 0,178206829 | 16,6319481740% |
| 0,4 | 4 | 0,277128129 | -0,411750149 | -0,118804553 | 0,118804553 | 0,292945596 | -5,3994555246% | 0,237609106 | 16,6319481576% |
| 0,5 | 4 | 0,346410162 | -0,514687687 | -0,148505691 | 0,148505691 | 0,366181995 | -5,3994554213% | 0,297011382 | 16,6319482656% |
| 0,6 | 4 | 0,415692194 | -0,617625224 | -0,178206829 | 0,178206829 | 0,439418394 | -5,3994554600% | 0,356413659 | 16,6319482394% |
| 0,7 | 4 | 0,484974226 | -0,720562761 | -0,207907967 | 0,207907968 | 0,512654794 | -5,3994554877% | 0,415815935 | 16,6319482207% |
| 0,8 | 4 | 0,554256258 | -0,823500298 | -0,237609106 | 0,237609106 | 0,585891193 | -5,3994554923% | 0,475218212 | 16,6319482067% |
| 0,9 | 4 | 0,623538291 | -0,926437836 | -0,267310245 | 0,267310243 | 0,659127591 | -5,3994554529% | 0,534620488 | 16,6319482436% |
| 1 | 4 | 0,692820323 | -1,029375373 | -0,297011382 | 0,297011382 | 0,732363991 | -5,3994554859% | 0,594022765 | 16,6319481870% |
| 2 | 4 | 1,385640646 | -2,058750746 | -0,594022765 | 0,594022765 | 1,464727982 | -5,3994554923% | 1,188045529 | 16,6319482067% |
| 3 | 4 | 2,078460969 | -3,088126119 | -0,891034147 | 0,891034147 | 2,197091972 | -5,3994554902% | 1,782068294 | 16,6319482001% |
| 4 | 4 | 2,771281292 | -4,117501492 | -1,188045530 | 1,188045530 | 2,929455963 | -5,3994554923% | 2,376091058 | 16,6319482067% |
| 5 | 4 | 3,464101615 | -5,146876865 | -1,485056911 | 1,485056911 | 3,661819954 | -5,3994554988% | 2,970113822 | 16,6319482263% |
| 6 | 4 | 4,156921938 | -6,176252238 | -1,782068294 | 1,782068294 | 4,394183944 | -5,3994554816% | 3,564136588 | 16,6319481740% |
| 7 | 4 | 4,849742261 | -7,205627611 | -2,079079676 | 2,079079676 | 5,126547935 | -5,3994554877% | 4,158159352 | 16,6319481927% |
| 8 | 4 | 5,542562584 | -8,235002984 | -2,376091058 | 2,376091058 | 5,858911926 | -5,3994554923% | 4,752182116 | 16,6319482067% |
| 9 | 4 | 6,235382907 | -9,264378357 | -2,67310244 | 2,67310244 | 6,591275917 | -5,3994554959% | 5,34620488 | 16,6319482176% |
| 10 | 4 | 6,92820323 | -10,29375373 | -2,970113823 | 2,970113823 | 7,323639907 | -5,3994554859% | 5,940227646 | 16,6319481870% |
| 20 | 4 | 13,85640646 | -20,58750746 | -5,940227645 | 5,940227645 | 14,64727982 | -5,3994554923% | 11,88045529 | 16,6319482067% |
| 30 | 4 | 20,78460969 | -30,88126119 | -8,910341468 | 8,910341468 | 21,97091972 | -5,3994554902% | 17,82068294 | 16,6319482001% |
| 40 | 4 | 27,71281292 | -41,17501492 | -11,88045529 | 11,88045529 | 29,29455963 | -5,3994554923% | 23,76091058 | 16,6319482067% |
| 50 | 4 | 34,64101615 | -51,46876865 | -14,85056911 | 14,85056911 | 36,61819954 | -5,3994554988% | 29,70113822 | 16,6319482263% |
| 60 | 4 | 41,56921938 | -61,76252238 | -17,82068294 | 17,82068294 | 43,94183944 | -5,3994554816% | 35,64136588 | 16,6319481740% |
| 70 | 4 | 48,49742261 | -72,05627611 | -20,79079676 | 20,79079676 | 51,26547935 | -5,3994554877% | 41,58159352 | 16,6319481927% |
| 80 | 4 | 55,42562584 | -82,35002984 | -23,76091058 | 23,76091058 | 58,58911926 | -5,3994554923% | 47,52182116 | 16,6319482067% |
| 90 | 4 | 62,35382907 | -92,64378357 | -26,7310244 | 26,7310244 | 65,91275917 | -5,3994554959% | 53,4620488 | 16,6319482176% |
| 100 | 4 | 69,2820323 | -102,9375373 | -29,70113823 | 29,70113823 | 73,23639907 | -5,3994554859% | 59,40227646 | 16,6319481870% |

| Sigma | n | G' = H' = = 2*sqrt(3)* *Sigma/(n+1) | E(X(1,n)) | E(X(2,n)) | E(X(3,n)) | G = = E(X(2,n)) - - E(X(1,n)) | (G'-G)/G | H = = E(X(3,n)) - - E(X(2,n)) | (H'-H)/H |
|---|---|---|---|---|---|---|---|---|---|
| 0,1 | 5 | 0,057735027 | -0,116296447 | -0,049501897 | 0 | 0,06679455 | -13,5632674025% | 0,049501897 | 16,6319483049% |
| 0,2 | 5 | 0,115470054 | -0,232592895 | -0,099003794 | 0 | 0,133589101 | -13,5632672666% | 0,099003794 | 16,6319481753% |
| 0,3 | 5 | 0,173205081 | -0,348889342 | -0,148505691 | 0 | 0,200383651 | -13,5632673162% | 0,148505691 | 16,6319482263% |
| 0,4 | 5 | 0,230940108 | -0,465185789 | -0,198007588 | 0 | 0,267178201 | -13,5632673054% | 0,198007588 | 16,6319482460% |
| 0,5 | 5 | 0,288675135 | -0,581482237 | -0,247509485 | 0 | 0,333972751 | -13,5632672213% | 0,247509485 | 16,6319481164% |
| 0,6 | 5 | 0,346410162 | -0,697778684 | -0,297011382 | 0 | 0,400767302 | -13,5632672731% | 0,297011382 | 16,6319482263% |
| 0,7 | 5 | 0,404145188 | -0,814075131 | -0,346513279 | 0 | 0,467561852 | -13,5632673100% | 0,346513279 | 16,6319483049% |
| 0,8 | 5 | 0,461880215 | -0,930371579 | -0,396015176 | 0 | 0,534356402 | -13,5632672731% | 0,396015176 | 16,6319482165% |
| 0,9 | 5 | 0,519615242 | -1,046668026 | -0,445517074 | 0 | 0,601150952 | -13,5632672156% | 0,445517074 | 16,6319481216% |
| 1 | 5 | 0,577350269 | -1,162964474 | -0,495018971 | 0 | 0,667945504 | -13,5632733378% | 0,495018971 | 16,6319481870% |
| 2 | 5 | 1,154700538 | -2,325928947 | -0,990037941 | 0 | 1,335891006 | -13,5632672795% | 0,990037941 | 16,6319481988% |
| 3 | 5 | 1,732050808 | -3,488893421 | -1,485056911 | 0 | 2,00383651 | -13,5632673162% | 1,485056911 | 16,6319482263% |
| 4 | 5 | 2,309401077 | -4,651857895 | -1,980075882 | 0 | 2,671782013 | -13,5632630054% | 1,980075882 | 16,6319481870% |
| 5 | 5 | 2,886751346 | -5,814822368 | -2,475094852 | 0 | 3,339727516 | -13,5632672990% | 2,475094852 | 16,6319482106% |
| 6 | 5 | 3,464101615 | -6,977786842 | -2,970113823 | 0 | 4,007673019 | -13,5632672946% | 2,970113823 | 16,6319481870% |
| 7 | 5 | 4,041451884 | -8,140751315 | -3,465132793 | 0 | 4,675618522 | -13,5632672916% | 3,465132793 | 16,6319482039% |
| 8 | 5 | 4,618802154 | -9,303715789 | -3,960151764 | 0 | 5,343564025 | -13,5632672893% | 3,960151764 | 16,6319481870% |
| 9 | 5 | 5,196152423 | -10,46668026 | -4,455170734 | 0 | 6,011509526 | -13,5632672587% | 4,455170734 | 16,6319482001% |
| 10 | 5 | 5,773502692 | -11,62964474 | -4,950189705 | 0 | 6,679455035 | -13,5632673378% | 4,950189705 | 16,6319481870% |
| 20 | 5 | 11,54700538 | -23,25928947 | -9,900379409 | 0 | 13,35891006 | -13,5632672795% | 9,900379409 | 16,6319481988% |
| 30 | 5 | 17,32050808 | -34,88893421 | -14,85056911 | 0 | 20,0383651 | -13,5632673162% | 14,85056911 | 16,6319482263% |
| 40 | 5 | 23,09401077 | -46,51857895 | -19,80075882 | 0 | 26,71782013 | -13,5632630054% | 19,80075882 | 16,6319481870% |
| 50 | 5 | 28,86751346 | -58,14822368 | -24,75094852 | 0 | 33,39727516 | -13,5632672990% | 24,75094852 | 16,6319482106% |
| 60 | 5 | 34,64101615 | -69,77786842 | -29,70113823 | 0 | 40,07673019 | -13,5632672946% | 29,70113823 | 16,6319481870% |
| 70 | 5 | 40,41451884 | -81,40751315 | -34,65132793 | 0 | 46,75618522 | -13,5632672916% | 34,65132793 | 16,6319482039% |
| 80 | 5 | 46,18802154 | -93,03715789 | -39,60151764 | 0 | 53,43564025 | -13,5632672893% | 39,60151764 | 16,6319481870% |
| 90 | 5 | 51,96152423 | -104,6668026 | -44,55170734 | 0 | 60,11509526 | -13,5632672587% | 44,55170734 | 16,6319482001% |
| 100 | 5 | 57,73502692 | -116,2964474 | -49,50189705 | 0 | 66,79455035 | -13,5632673378% | 49,50189705 | 16,6319481870% |

| Sigma | n | G' = H' = = 2*sqrt(3)* *Sigma/(n+1) | E(X(1,n)) | E(X(2,n)) | E(X(3,n)) | G = = E(X(2,n)) - - E(X(1,n)) | (G'-G)/G | H = = E(X(3,n)) - - E(X(2,n)) | (H'-H)/H |
|---|---|---|---|---|---|---|---|---|---|
| 0,1 | 6 | 0,049487166 | -0,126720636 | -0,064175504 | -0,020154683 | 0,062545132 | -20,8776700116% | 0,044020821 | 12,4176363246% |
| 0,2 | 6 | 0,098974332 | -0,253441272 | -0,128351008 | -0,040309367 | 0,125090264 | -20,8776698851% | 0,088041641 | 12,4176363885% |
| 0,3 | 6 | 0,148461498 | -0,380161908 | -0,192526512 | -0,06046405 | 0,187635397 | -20,8776699483% | 0,132062462 | 12,4176363757% |
| 0,4 | 6 | 0,197948664 | -0,506882544 | -0,256702016 | -0,080618733 | 0,250180529 | -20,8776699167% | 0,176083282 | 12,4176363502% |
| 0,5 | 6 | 0,24743583 | -0,63360318 | -0,32087752 | -0,100773417 | 0,312725661 | -20,8776698471% | 0,220104102 | 12,4176364523% |
| 0,6 | 6 | 0,296922996 | -0,760323816 | -0,385053023 | -0,1209281 | 0,375270793 | -20,8776699272% | 0,264124923 | 12,4176364438% |
| 0,7 | 6 | 0,346410162 | -0,887044452 | -0,449228527 | -0,141082783 | 0,437815925 | -20,8776699483% | 0,308145743 | 12,4176364742% |
| 0,8 | 6 | 0,395897327 | -1,013765088 | -0,513404031 | -0,161237467 | 0,500361057 | -20,8776698851% | 0,352166564 | 12,4176364651% |
| 0,9 | 6 | 0,445384493 | -1,140485725 | -0,577579535 | -0,181392151 | 0,56290619 | -20,8776699202% | 0,396187385 | 12,4176364013% |
| 1 | 6 | 0,494871659 | -1,267206361 | -0,641755039 | -0,201546834 | 0,625451322 | -20,8776699736% | 0,440208205 | 12,4176364013% |
| 2 | 6 | 0,989743319 | -2,534412721 | -1,283510078 | -0,403093668 | 1,250902643 | -20,8776698851% | 0,88041641 | 12,4176363502% |
| 3 | 6 | 1,484614978 | -3,801619082 | -1,925265116 | -0,604640501 | 1,876353966 | -20,8776699483% | 1,320624615 | 12,4176364353% |
| 4 | 6 | 1,979486637 | -5,068825442 | -2,567020155 | -0,806187335 | 2,501805287 | -20,8776699167% | 1,76083282 | 12,4176364140% |
| 5 | 6 | 2,474358297 | -6,336031803 | -3,208775194 | -1,007734169 | 3,127256609 | -20,8776699230% | 2,201041025 | 12,4176364013% |
| 6 | 6 | 2,969229956 | -7,603238164 | -3,850530233 | -1,209281003 | 3,752707931 | -20,8776699272% | 2,64124923 | 12,4176364013% |
| 7 | 6 | 3,464101615 | -8,870444524 | -4,492285271 | -1,410827837 | 4,378159253 | -20,8776699303% | 3,081457434 | 12,4176364377% |
| 8 | 6 | 3,958973274 | -10,13765088 | -5,13404031 | -1,61237467 | 5,00361057 | -20,8776698534% | 3,52166564 | 12,4176364013% |
| 9 | 6 | 4,453844934 | -11,40485725 | -5,775795349 | -1,813921504 | 5,629061901 | -20,8776699905% | 3,961873845 | 12,4176364013% |
| 10 | 6 | 4,948716593 | -12,67206361 | -6,417550388 | -2,015468338 | 6,254513222 | -20,8776699736% | 4,40208205 | 12,4176364013% |
| 20 | 6 | 9,897433186 | -25,34412721 | -12,83510078 | -4,030936676 | 12,50902643 | -20,8776698851% | 8,804164104 | 12,4176363502% |
| 30 | 6 | 14,84614978 | -38,01619082 | -19,25265116 | -6,046405014 | 18,76353966 | -20,8776699483% | 13,20624615 | 12,4176364353% |
| 40 | 6 | 19,79486637 | -50,68825442 | -25,67020155 | -8,061873352 | 25,01805287 | -20,8776699167% | 17,6083282 | 12,4176364140% |
| 50 | 6 | 24,74358297 | -63,36031803 | -32,08775194 | -10,07734169 | 31,27256609 | -20,8776699230% | 22,01041025 | 12,4176364013% |
| 60 | 6 | 29,69229956 | -76,03238164 | -38,50530233 | -12,09281003 | 37,52707931 | -20,8776699272% | 26,4124923 | 12,4176364013% |
| 70 | 6 | 34,64101615 | -88,70444524 | -44,92285271 | -14,10827837 | 43,78159253 | -20,8776699303% | 30,81457434 | 12,4176364377% |
| 80 | 6 | 39,58973274 | -101,3765088 | -51,3404031 | -16,1237467 | 50,0361057 | -20,8776698534% | 35,2166564 | 12,4176364013% |
| 90 | 6 | 44,53844934 | -114,0485725 | -57,75795349 | -18,13921504 | 56,29061901 | -20,8776699905% | 39,61873845 | 12,4176364013% |
| 100 | 6 | 49,48716593 | -126,7206361 | -64,17550388 | -20,15468338 | 62,54513222 | -20,8776699736% | 44,0208205 | 12,4176364013% |

| n | g(n) | h(n) |
|---|---|---|
| 2 | 1,1283791670 | n.a. |
| 3 | 0,8462843752 | 0,8462843752 |
| 4 | 0,7323639907 | 0,5940227646 |
| 5 | 0,6679455035 | 0,4950189705 |
| 6 | 0,6254513222 | 0,4402082050 |
| 7 | 0,5948041054 | 0,4046673114 |
| 8 | 0,5713754435 | 0,3794023676 |
| 9 | 0,5527157053 | 0,3603266738 |
| 10 | 0,5373956860 | 0,3452979396 |
| 11 | 0,5245198320 | 0,3330771153 |
| 12 | 0,5134954560 | 0,3228939849 |
| 13 | 0,5039129830 | 0,3142425616 |
| 14 | 0,4954792790 | 0,3067755711 |
| 15 | 0,4879783630 | 0,3002460517 |
| 16 | 0,4812471700 | 0,2944731270 |
| 17 | 0,4751599930 | 0,2893209990 |
| 18 | 0,4696181660 | 0,2846855300 |
| 19 | 0,4645430200 | 0,2804853930 |
| 20 | 0,4598709640 | 0,2766560440 |
| 21 | 0,4555499580 | 0,2731455450 |
| 22 | 0,4515369490 | 0,2699115720 |
| 23 | 0,4477959690 | 0,2669192510 |
| 24 | 0,4442967020 | 0,2641395560 |
| 25 | 0,4410133880 | 0,2615481170 |
| 26 | 0,4379239800 | 0,2591242940 |
| 27 | 0,4350094810 | 0,2568504840 |
| 28 | 0,4322534190 | 0,2547115690 |
| 29 | 0,4296414360 | 0,2526944780 |
| 30 | 0,4271609440 | 0,2507878530 |
| 31 | 0,4248008580 | 0,2489817650 |
| 32 | 0,4225513690 | 0,2472674960 |
| 33 | 0,4204037650 | 0,2456373480 |
| 34 | 0,4183502730 | 0,2440845050 |
| 35 | 0,4163839350 | 0,2426028980 |
| 36 | 0,4144984980 | 0,2411871100 |
| 37 | 0,4126883280 | 0,2398322850 |
| 38 | 0,4109483350 | 0,2385340540 |
| 39 | 0,4092738980 | 0,2372884800 |
| 40 | 0,4076608180 | 0,2360919970 |
| 41 | 0,4061052700 | 0,2349413720 |
| 42 | 0,4046037570 | 0,2338336620 |
| 43 | 0,4031530740 | 0,2327661880 |
| 44 | 0,4017502860 | 0,2317364940 |
| 45 | 0,4003926870 | 0,2307423340 |
| 46 | 0,3990777840 | 0,2297816480 |
| 47 | 0,3978032780 | 0,2288525360 |
| 48 | 0,3965670390 | 0,2279532480 |
| 49 | 0,3953670950 | 0,2270821700 |
| 50 | 0,3942016150 | 0,2262378060 |
| 51 | 0,3930688970 | 0,2254187700 |
| 52 | 0,3919673540 | 0,2246237780 |
| 53 | 0,3908955100 | 0,2238516320 |
| 54 | 0,3898519840 | 0,2231012200 |
| 55 | 0,3888354830 | 0,2223715030 |
| 56 | 0,3878448010 | 0,2216615120 |
| 57 | 0,3868788020 | 0,2209703410 |
| 58 | 0,3859364250 | 0,2202971400 |
| 59 | 0,3850166710 | 0,2196411130 |
| 60 | 0,3841185960 | 0,2190015150 |
| 61 | 0,3832413190 | 0,2183776430 |
| 62 | 0,3823840010 | 0,2177688400 |
| 63 | 0,3815458560 | 0,2171744810 |
| 64 | 0,3807261410 | 0,2165939810 |
| 65 | 0,3799241490 | 0,2160267900 |
| 66 | 0,3791392170 | 0,2154723830 |
| 67 | 0,3783707120 | 0,2149302680 |
| 68 | 0,3776180370 | 0,2143999770 |
| 69 | 0,3768806230 | 0,2138810700 |
| 70 | 0,3761579330 | 0,2133731270 |
| 71 | 0,3754494530 | 0,2128757490 |
| 72 | 0,3747546980 | 0,2123885600 |
| 73 | 0,3740732020 | 0,2119112010 |
| 74 | 0,3734045240 | 0,2114433320 |
| 75 | 0,3727482440 | 0,2109846280 |
| 76 | 0,3721039610 | 0,2105347790 |
| 77 | 0,3714712890 | 0,2100934920 |
| 78 | 0,3708498640 | 0,2096604860 |
| 79 | 0,3702393370 | 0,2092354930 |
| 80 | 0,3696393720 | 0,2088182580 |
| 81 | 0,3690496510 | 0,2084085360 |
| 82 | 0,3684698670 | 0,2080060940 |
| 83 | 0,3678997260 | 0,2076107080 |
| 84 | 0,3673389480 | 0,2072221650 |
| 85 | 0,3667872630 | 0,2068402600 |
| 86 | 0,3662444120 | 0,2064647970 |
| 87 | 0,3657101470 | 0,2060955900 |
| 88 | 0,3651842300 | 0,2057324550 |
| 89 | 0,3646664320 | 0,2053752220 |
| 90 | 0,3641565330 | 0,2050237240 |
| 91 | 0,3636543210 | 0,2046778020 |
| 92 | 0,3631595930 | 0,2043373010 |
| 93 | 0,3626721520 | 0,2040020770 |
| 94 | 0,3621918100 | 0,2036719850 |
| 95 | 0,3617183850 | 0,2033468900 |
| 96 | 0,3612517010 | 0,2030266610 |
| 97 | 0,3607915900 | 0,2027111720 |
| 98 | 0,3603378890 | 0,2024003010 |
| 99 | 0,3598904410 | 0,2020939290 |
| 100 | 0,3594490920 | 0,2017919450 |